\documentclass[11pt]{article}

\usepackage[utf8]{inputenc}
\usepackage[T1]{fontenc}
\usepackage{amsmath}
\usepackage{amsfonts}
\usepackage{amssymb}
\usepackage{float}
\usepackage[version=4]{mhchem}
\usepackage{stmaryrd}
\usepackage{graphicx}
\usepackage{array, xcolor, lipsum, bibentry, fancyhdr}
\usepackage[export]{adjustbox}
\graphicspath{ {images/} }
\usepackage{hyperref}

\hypersetup{
    colorlinks,
    linkcolor={red!50!red},
    citecolor={blue!50!blue},
    urlcolor={blue!80!blue}
}

\newcounter{susis}

\title{Pricing Multi-strike Quanto Call Options on Multiple Assets with Stochastic Volatility, Correlation, and Exchange Rates}
\author{
  {\it Boris Ter-Avanesov$^{1}$ and Gunter Meissner$^{2}$} \\
  \\
  \href{https://www.columbia.edu}{Columbia University}, New York, USA \\
  $^1$ {\href{mailto:bt2522@columbia.edu}{\small\texttt{bt2522@columbia.edu}}}, $^2$ {\href{mailto:gm2718@columbia.edu}{\small\texttt{gm2718@columbia.edu}}}
}
\date{\today}

\begin{document}

\maketitle

\begin{abstract}
Quanto options allow the buyer to exchange the foreign currency payoff into the domestic currency at a fixed exchange rate. We investigate quanto options with multiple underlying assets valued in different foreign currencies each with a different strike price in the payoff function. We carry out a comparative performance analysis of different stochastic volatility (SV), stochastic correlation (SC), and stochastic exchange rate (SER) models to determine the best combination of these models for Monte Carlo (MC) simulation pricing. In addition, we test the performance of all model variants with constant correlation as a benchmark. We find that a combination of GARCH-Jump SV, Weibull SC, and Ornstein Uhlenbeck (OU) SER performs best. In addition, we analyze different discretization schemes and their results. In our simulations, the Milstein scheme yields the best balance between execution times and lower standard deviations of price estimates. Furthermore, we find that incorporating mean reversion into stochastic correlation and stochastic FX rate modeling is beneficial for MC simulation pricing. We improve the accuracy of our simulations by implementing antithetic variates variance reduction. Finally, we derive the correlation risk parameters Cora and Gora in our framework so that correlation hedging of quanto options can be performed.
\\

\end{abstract}

\\

\noindent This paper is structured as follows: Section 1 provides the introduction and outlines the methodology. Section 2 gives a detailed description of the different stochastic differential equation (SDE) models used for volatility, correlation, and exchange rates. Section 3 focuses on the options being studied, with an emphasis on the payoff structures, underlying assets, and the overall model framework. Section 4 discusses the three discretization schemes, including their adaptation for SDEs with Jumps, and the Monte Carlo simulation pricing methodology with antithetic variates for variance reduction. Section 5 presents a comparison of results, accompanied by a discussion of output plots and tables. Section 6 delves into the derivation of the Cora and Gora correlation risk parameters. Section 7 concludes the paper, followed by a brief outline of future work, the bibliography and references, and a list of figures.

\section{INTRODUCTION \& METHOD}
In modern finance, tradable assets are typically modeled with stochastic volatility, which was introduced by Hull and White in 1987 \cite{Hull1987}. Stein and Stein (1991) introduced a mean-reverting stochastic volatility model where the volatility of asset returns follows a Brownian motion mean-reverting process \cite{Stein1991}. This model reflects the observation that volatility tends to return to a long-term average level over time, providing a more realistic depiction of market behavior compared to constant volatility models. In 1993, Heston \cite{Heston1993} correlated the stochastic stock price and stochastic stock price volatility by correlating their Brownian motions. Another example of a stochastic volatility model was developed by Ball and Roma (1994) \cite{Ball1994}. Their model presents a framework for option pricing that accounts for stochastic volatility by simplifying the Fourier option pricing techniques and implementing power series methods. They demonstrate that the characteristic function of the average variance is crucial in this approach, particularly when there is no correlation between security price innovations and volatility. This model corrects certain biases in the Black-Scholes model, improving on Stein and Stein's analysis \cite{Ball1994}. Bates' (1996) \cite{Bates1996} model further extends Heston’s model by incorporating jumps in the asset price process, thereby capturing sudden, large movements in the market, which is a common feature observed in financial time series data. Modeling volatility as stochastic captures the empirical observation that market volatility tends to cluster over time, reflecting periods of high and low market uncertainty, which cannot be explained by a constant volatility model. Moreover, the phenomenon of volatility smiles has been studied extensively and seems to be alleviated by models with non-constant volatility (\cite{Heston1993}, \cite{Vagnani2009}, \cite{Sheraz2013}, \cite{Hamza2006}, \cite{Marcato2018}). 
\\

However, it is much less common in comparison to see such models extended further with a correlation that varies stochastically over time. Some existing research concerned with modeling correlations as stochastic is Engle 2002 \cite{Engle2002}, Lu \& Meissner 2014 \cite{LuMeissner2020}, Buraschi et al. 2010 \cite{Buraschi2010} \cite{BuraschiPortfolio2010}, and Da Fonseca et al. 2007/2008 \cite{Fonseca2007}. Modeling correlation as stochastic (SC) is beneficial because it reflects the reality that correlations between asset returns are not constant and can change due to varying market conditions, such as shifts in economic cycles or changes in investor sentiment. This variability in correlation can significantly impact the pricing and hedging of multi-asset derivatives. In his research, \textit{Pricing Foreign Equity Options with Stochastic Correlation and Volatility} (2009), Jun Ma develops a novel model of this type for foreign equity option pricing. Foreign equity, FX, and currency derivatives are widely traded on a global scale. Crucially, participants incur additional risk due to exchange rate uncertainty when trading foreign equity options, as highlighted by Ma in his paper \cite{Ma2009}. Moreover, when trading derivatives that rely on multiple underlying assets, participants also incur an additional correlation risk that has to be accounted for. As clarified by Ma, Quanto options do not yield to pricing via the BS risk-neutral framework when we incorporate stochastic correlation \cite{Ma2009}. In cases when a simple closed-form solution is unknown, some popular alternatives for pricing such derivatives are numerical methods, simulations, or series solutions. 
\\

The primary aim of this paper is to use simulations to tackle the problem of pricing Quanto options on two and three underlying assets under stochastic correlation and volatility driven by different stochastic differential equations (SDEs). The following models are tested and compared: Heston, GARCH, GARCH-Jump, 3/2 diffusion, and Bates for volatility, and Jacobi, Wright-Fisher diffusion, Weibull diffusion, and a mean-reverting SC for correlation. The study is focused specifically on Quanto options on two or three foreign equity market indices. These options act like a basket correlation option with the payoff depending on multiple correlated assets but also on exchange rates between the currencies of the indices. We test three different models of exchange rate dynamics, with both rates being either GBM, a mean reverting SDE inspired by the OU process, or an exponential levy process that incorporates jumps. The stochastic differential equations governing all of these SV, SC, and SER models and their key features can be found in section 2 of the paper. The most unique feature of Quanto options is the payoff structure since it is paid in the foreign currency of the underlying but then converted to the domestic currency. Details of the payoff structure and the overall model outlines are discussed fully in section 3 of the paper. Section 4 of our paper outlines the methodology of the Monte Carlo simulation and clarifies details of the discretization schemes for the SDEs, whilst section 5 discusses the results. Due to the additional correlation risk, it is prudent to also consider how to effectively hedge such products with Cora and Gora, which is done in section 6.
\\

We collect observed market prices for the indices SP500, FTSE100, and STOXX600 (underlying assets), as well as the GBP/USD, EUR/USD, and EUR/GBP exchange rates from Yahoo Finance. The code written for this paper allows the user to select what date range the option will be over and, hence, what data to collect. For the paper, we focus on 2021-2022 and 2022-2023 as two specific time periods to test. We perform the comparison of all model variants and the discretization schemes for both cases of the option and for both of these date ranges as the lifetimes of the option. The constant parameters of the SDEs of the volatilities, correlations, and exchange rates are calibrated to real market data based on summary statistics of the SP500, FTSE100, and STOXX600 indices and USD/GBP, USD/EUR exchange rates values. For example, aside from the starting points of all the processes, we also selected the volatility parameters of the FX rate SDEs based on the rolling standard deviations for the exchange rates. Similarly, the mean-reversion level (\(\mu\)) and rate (\(\theta\)) of the Ornstein-Uhlenbeck (OU) process for modeling FX rates are determined using historical average exchange rates and autocorrelation analysis. Likewise, we use historical long-run averages, rolling standard deviations, and rolling correlations to select the parameters of the SV and SC SDEs. We ensure that look-ahead bias is avoided by only using data available on the start date of each option for the calibration. To facilitate a visual comparison with the plots of paths of different processes produced by the simulation, we standardize the plots of the observed trajectories of our underlying assets and exchange rates in different years to plot them with the same starting points with 100 used as an example value (Figures 29 and 30). We also plot rolling window volatilities and correlations of the assets and exchange rates for different window sizes (Figures 31 and 32). We use the 13-week US Treasury Bill rate from Yahoo Finance as the US domestic interest rate (Figure 24). The Bank of England \cite{BoE} and European Central Bank \cite{ECB} interest rates are used for the two foreign interest rates. 
\\

Figure 1 shows the starting values for the underlying assets and the exchange rates. These values are used as the starting parameters for the MC simulations as discussed in section 4. 

\begin{figure}[h]
    \centering
    \includegraphics[width=\linewidth]{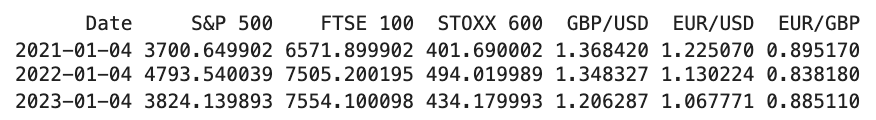} % 
    \caption{Values of indices and exchange rates.}
    \label{fig:table_values}
\end{figure}

\section{MODELS of STOCHASTIC VOLATILITY, CORRELATION \& EXCHANGE RATES}
The quest for pricing models that incorporate random volatility is driven by empirical evidence from various studies of financial time series supporting the hypothesis of stochastic volatility. Ma (2009) emphasizes that implied volatilities, calculated using the Black-Scholes formula \cite{Black1973}, exhibit random fluctuations over time, manifesting in the term structure of implied volatility. Moreover, when we fix a maturity time T, and consider varying strike price K, it can be seen that implied volatility is higher for options contracts with only the moneyness level being different, which is often referred to as the volatility smile or skew. This pattern arises because the Black-Scholes model assumes constant volatility, which does not align with its market behavior, where volatility varies over time and across strike prices. While the volatility smile is typically symmetric for both puts and calls, this pattern is more common in currency (FX) and commodity markets, where implied volatility curves form a valley or smile shape, rising at both ends for deep in-the-money (ITM) and OTM options. In contrast, equity markets often exhibit a downward-sloping implied volatility graph, commonly referred to as a skew or smirk. In such markets, OTM put options tend to have higher implied volatility than ITM put options (left to right downward slope) and OTM call options tend to have lower implied volatility than ITM call options (right to left downward slope). Also, this pattern is more pronounced for puts than for calls due to increased demand for downside protection, reflecting perceived risks of large negative movements in asset prices. According to Ma, stochastic volatility models elucidate deviations from constant implied volatility, and it has been shown that they can capture the volatility smile better. 
\\

Traditionally, in the vast body of financial and economic literature on multi-asset options, the correlation coefficient between correlated variables has been assumed to be constant (e.g., see Black and Scholes, 1973 \cite{Black1973}; Margrabe, 1978 \cite{Margrabe1978}; Garman, 1992 \cite{Garman1992}). However, Ma (2009) highlights that relying on long-term estimates of constant correlation can be misleading, potentially resulting in significant mispricing and risk management issues. Historical correlations must be used with caution as they can be more unstable than volatility \cite{Ma2009}. An alternative approach involves inferring implied correlations from market prices, akin to implied volatility, which offers an estimation of stochastic correlation based on market data \cite{Ma2009}. In this study, we test different models of stochastic volatility of the underlying assets, of stochastic correlation between their Brownian motions, and of stochastic exchange rates. 

\subsection{Volatility Models}
As mentioned in the introduction, we test five models for stochastic volatility: Heston, 3/2 volatility, GARCH, Bates, and GARCH-Jump. The first two rely on a mean reversion drift term, with the parameters kappa and theta being the rate of mean reversion and the long-run value to which the volatility process reverts, respectively. Also, the first two have the parameter sigma to control the standard deviation of the random fluctuations. 3/2 volatility can be thought of as a higher-order extension of Heston volatility. Bates and GARCH-Jump processes introduce jumps to replicate the behavior of sudden moves of volatility observed in markets with Poisson-process-driven, Normally distributed jumps. We ensure that the simulation functions do not produce negative volatility for all of the SDEs of stochastic volatility models.

\subsubsection{Heston Model}
\begin{equation}
dv_t = \kappa (\theta - v_t) dt + \sigma \sqrt{v_t} dW_t
\end{equation}

Note that setting \begin{equation} 2\kappa\theta > \sigma^2 \end{equation} in the Heston model ensures the process is strictly positive (Feller condition \cite{Albrecher2007}).  
 
\subsubsection{GARCH Inspired Model}
The time-homogeneous GARCH process satisfies the following linear SDE according to Li et al. as described in their 2018 paper \cite{LiMercurio2018}. Hence, this model is very similar to the Heston model discussed above, but the diffusion term is multiplied by $\sqrt{v_t}$ to raise the power on the volatility component from 1/2 to 1 ... 

\begin{equation}
dv_t = \kappa (\theta - v_t) dt + \sigma v_t dW_t
\end{equation}

\subsubsection{GARCH Inspired Model with Jumps}
The GARCH-Jump model extends the standard GARCH (Generalized Autoregressive Conditional Heteroskedasticity) framework by incorporating jumps, providing a more comprehensive tool to capture the aforementioned volatility dynamics observed in financial markets. This model is particularly useful in capturing the sudden large movements or jumps in asset prices that cannot be explained by continuous processes alone. The GARCH-Jump model was developed and extensively analyzed by Duan et al. (2004), who highlighted its efficacy in better fitting historical time series data and explaining the observed volatility smile in option prices. Their research demonstrated that incorporating jumps into the GARCH framework significantly improves the model's performance in capturing the empirical features of asset returns and volatility \cite{Duan2004}. We implement a variant of GARCH-Jump SV by adding a jump term to our GARCH-inspired model discussed above. 

\begin{equation}
dv_t = \kappa (\theta - v_t) dt + \sigma v_t dW_t + \zeta dJ_t
\end{equation}

\noindent In our code implementation, the jumps are modeled as a compound Poisson process where the jump sizes are normally distributed. Specifically, \( dJ_t \) is constructed by first generating a number of jumps using a Poisson distribution with intensity parameter \( \lambda \). For each jump, the size is drawn from a normal distribution with mean \( \mu_J \) and standard deviation \( \sigma_J \). The total jump impact \( dJ_t \) is then the sum of all individual jump sizes occurring within a given time interval \( dt \). Hence, the jumps are modeled as follows: 

\begin{equation}
dJ_t = \sum_{i=1}^{N_t} Y_i \sim \text{Poisson}(\lambda \cdot dt) \times \mathcal{N}(\mu_J, \sigma_J^2)
\end{equation}

\noindent Here, \( N(t) \) is a Poisson process with intensity \( \lambda \), and \( Y_i \) are i.i.d. normal random variables with mean \( \mu_J \) and variance \( \sigma_J^2 \). Duan et al. (2004) employed the GARCH-Jump model to explore option pricing under conditions where both price and volatility exhibit jump-diffusion behavior. They found that the GARCH-Jump model provides a robust framework for understanding and predicting market behaviors characterized by sudden and significant changes. The model's ability to capture jumps makes it particularly valuable for pricing derivatives and managing financial risk in environments subject to abrupt market movements \cite{Duan2004}.

\subsubsection{Bates Model}
The Bates volatility model extends Heston volatility with jumps \cite{Bates1996}. Here, the parameters kappa, theta, and sigma have the same use as for the Heston volatility model. 

\begin{equation}
dv_t = \kappa (\theta - v_t) dt + \sigma \sqrt{v_t} dW_t + \zeta dJ_t
\end{equation}
where \( dJ_t \) represents the jumps with normally distributed jump sizes. \(mu_J\) and \(sigma_J\) are two more parameters in the simulation of this model, which control the mean size and standard deviation of the jumps, respectively, as discussed above. Here \( \zeta \) is a multiplier that controls the magnitude of the effect of the jumps as above. 

\subsubsection{3/2 Model}
The 3/2 stochastic volatility model is an extension of the constant elasticity of variance (CEV) model and was developed by Carr and Sun (2007) to better capture the dynamics of financial markets. In this model, the volatility of the underlying asset is driven by a process that is proportional to the power 3/2 of the volatility itself. The stochastic differential equation (SDE) governing the 3/2 model is given by:
\begin{equation}
dv_t = (\omega - \theta v_t) v_t dt + \sigma v_t^{3/2} dW_t,
\end{equation}

In this context:
\begin{itemize}
    \item \( \omega \) is the speed of mean reversion, determining how quickly the process reverts to its long-term mean \( \theta \).
    \item \( \theta \) is the long-term mean level of the variance process.
    \item \( \sigma \) represents the volatility of the variance process, indicating the magnitude of random fluctuations.
\end{itemize}

\noindent The 3/2 model is particularly useful in capturing the empirical features of volatility observed in financial markets, such as the leverage effect and the fact that volatility tends to spike during market downturns \cite{Carr2007}. Carr and Sun (2007) developed this model to provide a more accurate framework for pricing options and other derivative securities. According to Carr and Sun, the 3/2 model has several desirable properties. The process remains non-negative and exhibits mean-reverting behavior, where the speed of mean reversion is proportional to the level of the process. The 3/2 model also yields closed-form solutions for the joint Fourier-Laplace transform of returns and their quadratic variation, which is useful for efficiently pricing and hedging derivatives \cite{Carr2007}.

\subsection{Correlation Models}
Correlations can be influenced by factors such as industrial production, T-bill rates, and unanticipated inflation, often acting as a business cycle indicator. Even after adjusting for business cycle effects, correlation risk persists (Driessen, Maenhout, and Vilkov, 2006 \cite{Driessen2006}). Although the correlation coefficient between two assets is not directly tradable, it remains crucial to devise hedging strategies for correlation risk. Developing robust frameworks for constructing portfolios to hedge against correlation risk can ensure more secure risk management practices \cite{Ma2009}. For all of the stochastic correlation models, the SDEs have mean reversion or bounds or are clipped to ensure that correlations remain within [-1, 1]. Four models for stochastic correlation of increasing complexity are implemented in this study. The simplest is a stochastic correlation SDE inspired by the modeling of processes in studies of genetics, which has a diffusion term, making sure it stays within [-1, 1]. The next simplest is the Jacobi correlation used by Ma, which gives the user the option to keep the correlation process within bounds h and f by altering this diffusion term. The second most complicated model is the mean-reverting extension of the first two simpler models. The most complex model that is tested is the Weibull distribution stochastic correlation model. 

\subsubsection{Wright-Fisher (WF) Model}
Wright-Fisher diffusions are used in biology and biochemistry to model gene frequencies and other natural/bodily processes. As explained in their paper, \textit{A mean-reverting SDE on correlation matrices}, Ahdida et al focus on stochastic differential equations 'valued on correlation matrices' \cite{Ahdida2013} and develop a mean-reverting extension of the Wright-Fisher SDE to model correlations in finance. In our paper, we test both the original diffusion on [-1,1] and the mean-reverting extension developed by Ahdida et al. as alternatives for modeling correlation. We do not implement the original version of the diffusion on [0, 1] since we want the process to mimic the variability across the range [-1, 1] when modeling financial correlations. The classic WF diffusion on [-1,1] has the SDE: 

\begin{equation}
d\rho_t = \kappa (\bar{\rho} - \rho_t) dt + \sigma \sqrt{1 - \rho_t^2} dW_t
\end{equation}

\noindent Here, $\rho_t$ represents the instantaneous correlation at time $t$. $\kappa$ is the mean reversion rate, determining how quickly the process reverts to the long-term mean correlation $\bar{\rho}$. $\bar{\rho}$ is the long-term mean correlation towards which $\rho_t$ reverts. $\sigma$ represents the volatility of the correlation process, indicating the extent of random fluctuations around the mean, and $dW_t$ is a standard Wiener process, as usual. Ahdida et al. highlight that the term \( \sqrt{1 - \rho_t^2} \) ensures that the correlation \( \rho_t \) remains within the interval \([-1, 1]\). This term becomes zero when \( \rho_t \) approaches the boundaries, preventing it from exceeding these limits. This bounded characteristic makes this version of the WF diffusion particularly suitable for modeling correlations in finance, where it is critical to maintain realistic correlation values \cite{Ahdida2013}.

\subsubsection{Jacobi Process}
The Jacobi process is used to model stochastic correlation and is described by the following stochastic differential equation (SDE):
\begin{equation}
d\rho_t = \kappa (\bar{\rho} - \rho_t) dt + \sigma \sqrt{(h - \rho_t)(\rho_t - f)} dW_t
\end{equation}
where \( \rho_t \) represents the correlation at time \( t \), \( \kappa \) is the mean reversion rate, $\bar{\rho}$  is the long-term mean correlation, and \( h \) and \( f \) are the upper and lower bounds of the correlation, respectively. The term \( dW_t \) denotes a standard Wiener process as usual. The parameter \( \kappa \) determines the speed at which the correlation reverts to its long-term mean $\bar{\rho}$. A higher value of \( \kappa \) implies a faster reversion. The parameters \( h \) and \( f \) set the natural boundaries for the correlation, ensuring that it remains within a realistic range. The volatility parameter \( \sigma \) controls the amplitude of fluctuations around the mean \cite{Ma2009}.
\\

The Jacobi process is particularly useful for modeling correlation because it can capture both the mean-reverting nature and the bounded behavior of correlation coefficients. The square root term \( \sqrt{(h - \rho_t)(\rho_t - f)} \) ensures that the correlation stays within the interval \( (f, h) \). Ma (2009) introduced the use of the Jacobi process in the context of pricing foreign equity options with stochastic correlation. This model allows for a more accurate reflection of market dynamics compared to constant correlation models. The Jacobi process can be seen as an extension of other mean-reverting processes, such as the Ornstein-Uhlenbeck process, but with the added complexity of bounded behavior. This makes it particularly suited for financial applications since correlations tend to naturally exhibit such characteristics \cite{Ma2009}.

\subsubsection{Mean-Reverting Correlation}
This is an extension of the WF diffusion on [-1, 1] developed by Ahdida et al. (Ahdida 2013), which includes an adaptive mean-reversion component. The mean-reverting SDE is given by:

\begin{equation}
d\rho_t = (\kappa (\bar{\rho} - \rho_t) - \sigma^2 \rho_t) dt + \sigma \sqrt{1 - \rho_t^2} dW_t
\end{equation}

\noindent Here \( \rho_t \), \( \kappa \), \(bar{\rho}\),  \( \sigma \), are all the same as for the WF correlation model discussed above. Also, the term \( \sqrt{1 - \rho_t^2} \) has the same use of ensuring that the correlation \( \rho_t \) remains within the interval \([-1, 1]\). By adjusting the mean-reverting drift term \(\kappa (\bar{\rho} - \rho_t) - \sigma^2 \rho_t\), Ahdida et al extend the process to capture the empirically observed tendency of correlations to revert to a long-term mean with the inclusion of the \(\sigma^2 \rho_t\) to ensure that the reversion speed adjusts dynamically based on the current correlation level \cite{Ahdida2013}.

\subsubsection{Weibull Model}
The Weibull model is characterized by the following stochastic differential equation:
\begin{equation}
d\rho_t = -\alpha (\rho_t - \mu_W) dt + b_1 b_2 dW_t
\end{equation}
where \( \mu_W \) is a mean term derived from Weibull distribution parameters. This model was originally proposed by Miñano et al. (2013) for wind speed modeling, demonstrating its efficacy in generating wind speed trajectories with desired statistical properties \cite{Minano2013}. In this context, \( b_1 \) and \( b_2 \) are diffusion terms uniquely defined to ensure the model captures the desired statistical properties. Specifically, \( b_1 \) and \( b_2 \) are given by:
\begin{align}
b_1(\rho_t) &= \frac{2 \alpha}{p_W(\rho_t)}, \\
b_2(\rho_t) &= \lambda \Gamma\left(1 + \frac{1}{k}, \left(\frac{\rho_t}{\lambda}\right)^k\right) - \mu_W e^{-(\rho_t/\lambda)^k},
\end{align}
where \( p_W(\rho_t) \) is the probability density function (PDF) of the Weibull distribution, \( \lambda \) is the scale parameter, \( k \) is the shape parameter, and \( \Gamma \) represents the Gamma function. The inclusion of \( b_1 \) and \( b_2 \) ensures the model's diffusion term is appropriately scaled to reflect the nature of the underlying process. This formulation makes the Weibull model particularly suitable for phenomena where the Weibull distribution provides a good fit, such as wind speed data and potentially skewed financial data. Specifically, this model was designed to simulate wind speeds that follow a Weibull distribution and exhibit exponential autocorrelation, as discussed in the work by Miñano et al. \cite{Minano2013}. Their objective was to accurately replicate the statistical properties of wind speed for applications in power systems, highlighting the model's ability to generate realistic wind speed trajectories for various simulations and analyses. To ensure the mathematical validity of the model, it is essential that $ \rho_t $ paths stay non-negative and $ k > 0 $, as the Weibull distribution is defined only for non-negative values and requires a positive shape parameter. These conditions guarantee that the diffusion terms \( b_1 \) and \( b_2 \) are well-defined, ensuring the stochastic process remains within the domain where the Weibull distribution accurately describes the behavior. Therefore, this SC model works in situations when we want to model positive correlations. 
\\

It has been shown that the Johnson SB distribution is a best-fit distribution for equity and default probability correlation distributions \cite{Kaplan2018}. Also, Gunter Meissner et al conducted a comprehensive analysis of correlation behavior based on daily closing prices of 30 stocks within the Dow Jones index, spanning the period from January 1972 to October 2012 \cite{Meissner2019}. Their findings indicate that correlation levels are at their lowest during periods of robust economic growth, wherein equity prices are predominantly influenced by idiosyncratic factors rather than broader macroeconomic conditions. Conversely, during recessions, correlation levels typically rise as macroeconomic factors overshadow idiosyncratic influences, leading to simultaneous downturns across multiple stocks. Furthermore, the volatility of correlations is observed to be lowest during economic expansions and higher during normal periods and recessions \cite{Meissner2019}. The study also highlighted that the Johnson SB distribution, characterized by its shape, location, and scale parameters, provided the most accurate fit for modeling these correlations. This distribution's flexibility effectively captures the intricate properties and variations of the correlation data observed under different economic conditions \cite{Meissner2019}. Something we want to explore further in future work is to apply the approach used for the Weibull correlation model to capture statistical properties of the distribution in the SDE to develop a stochastic correlation model that captures desired properties of the Johnson SB distribution and incorporates mean reversion. 
\\

\subsection{Models of Stochastic Exchange Rates}
The domestic currency for the investor in these Quanto options is US dollars. In our paper, one of the underlying assets is denominated in the home currency (US dollar). For all choices of correlation and volatility, the underlying assets are modeled with the same SDEs throughout. We test three SDEs for the exchange rates (scenarios 1-3). The simplest model is for both exchange rates to be GBM \cite{Bachelier1900} with different parameters based on observed starting values and summary statistics (as outlined in section 1). We also test a model with mean-reversion incorporated into the SDEs used for both exchange rates. It has been shown in many studies that foreign exchange rates tend to exhibit strong mean reversion \cite{Jorion1996} \cite{Taylor2001}. In the third model, both exchange rates follow an exponential Lévy process, which incorporates jumps into the GBM model of scenario 1 \cite{Tankov} \cite{Andersen2012}. The foreign risk-free interest rate is \( r_{f_1} \). This is the risk-free rate of return in the first foreign currency, which could be the interest rate of a government bond or another risk-free security in the foreign market of GBP. \( r_{f_2} \) is the risk-free rate of return in the second foreign currency, and \( r_d \) is the domestic interest rate. In the equations below, \( \theta \) represents the rate of mean reversion, \( \mu \) is the mean level to which the process reverts, \( \sigma_{FX} \) is the volatility, and \( dW\textsubscript{FX}(t) \) is the increment of the Wiener process.

\subsubsection{Geometric Brownian Motion (GBM)}
The GBM is characterized by its exponential growth with constant drift and volatility \cite{Bachelier1900}. 

\begin{equation}
dFX(t) = FX(t) \left( (r_f - r_d - 0.5 \sigma_{FX}^2) dt + \sigma_{FX} dW\textsubscript{FX}(t) \right)
\end{equation}

\subsubsection{Ornstein-Uhlenbeck (OU) Process}
In this scenario, we model the exchange rates as an OU process, which introduces mean reversion into the drift term.

\begin{equation}
\begin{aligned}
dFX(t) &= \theta (\mu - FX(t)) dt + FX(t) \left( (r_f - r_d - 0.5 \sigma_{FX}^2) dt + \sigma_{FX} dW\textsubscript{FX}(t) \right)
\end{aligned}
\end{equation}

\subsubsection{Exponential Lévy Process}
In this scenario, the exchange rates follow an exponential Lévy process, which incorporates jumps into the SDEs, accounting for sudden and significant changes in the exchange rates \cite{Tankov} \cite{Hakansson2015}. The key parameters in this model include \( \sigma_{FX} \) for volatility, \( \lambda_L \) for the jump intensity, \( \mu_L \) for the mean jump size, and \( \sigma_L \) for the jump size volatility.

\begin{equation}
\begin{aligned}
dFX(t) &= FX(t) \left( (r_f - r_d - 0.5 \sigma_{FX}^2) dt + \sigma_{FX} dW\textsubscript{FX}(t) \right) + dJ_t
\end{aligned}
\end{equation}

\noindent Where \( dJ_t \) is the jump component, modeled as before (sections 2.1.3 and 2.1.4) for the GARCH-Jump and Bates SV. These jumps account for sudden, significant changes in the exchange rates, making the exponential Lévy process a suitable candidate for modeling FX rates with jumps \cite{Hakansson2015} \cite{Christensen2013}.

\section{BASKET QUANTO CALLS}
We test all combinations of choices of the 5 SV, 4 SC, and 3 SER models outlined above for pricing 2 types of basket Quanto call options. Case 1 involves two underlying assets and one exchange rate, whilst Case 2 has three underlying assets and two exchange rates. Both Cases 1 and 2 include an underlying asset in the domestic currency of US dollars, and the other asset(s) must be converted into dollars using the exchange rate(s). In this paper, we model the domestic and foreign interest rates as constants throughout the lifetime of the options. We pick the values observed on the start dates of the options. However, as shown in Figure 24, the interest rates are not usually constant in practice, even over a 1-year time window. We use the 13-week US Treasury Bills rate from Yahoo Finance as the US domestic interest rate, and it can be seen to change in value significantly throughout 2022-2023 (more than 5\% change) but remain relatively constant in the prior year. Similar observations can be made with the BoE rates. Thus, it could be prudent to model interest rates as a deterministic function of time, mixed jump diffusion, or discrete event/jump process (compound Poisson or Hawkes, for example), with appropriate adjustments for monthly or quarterly frequencies but this extension is left for future work. In addition, modeling interest rates with SDEs would also allow us to explore whether we should introduce correlations between them. As seen in Figure 24, all three interest rates appear to be strongly positively correlated. 
\\

The overall models for the pricing of our Quanto calls using specific choices of the SDE models for SV, SC, and SER can be written as follows: 

\subsection{Case 1: Two Underlying Assets - Single FX Rate}
\subsubsection{Variables}
\begin{itemize}
    \item \( S_{GBP}(t) \): Price of the 1st underlying (e.g. GBP asset).
    \item \( S_{USD}(t) \): Price of the 2nd underlying (e.g. USD asset).
    \item \( FX_{GBP}(t) \): FX rate (e.g. for USD/GBP).
\end{itemize}

\subsubsection{Stochastic Differential Equations}

\paragraph{Underlying Asset Prices}
\begin{equation}
dS_{USD}(t) = S_{USD}(t) \left( r_d dt + \sqrt{v_{USD}(t)} dW_{USD}(t) \right)
\end{equation}

\begin{equation}
dS_{GBP}(t) = S_{GBP}(t) \left( r_{f_1} dt + \sqrt{v_{GBP}(t)} dW_{GBP}(t) \right)
\end{equation}

\paragraph{SV Choice (same process for both underlying assets)}

\begin{equation}
dv\textsubscript{USD}_t = m\textsuperscript{v}(v\textsubscript{USD}_t, t) dt + s\textsuperscript{v}(v\textsubscript{USD}_t, t) dW\textsuperscript{v\_USD}_t + \omega\textsuperscript{v}(v\textsubscript{USD}_t, t) dJ\textsuperscript{v\_USD}_t
\end{equation}

\begin{equation}
dv\textsubscript{GBP}_t = m\textsuperscript{v}(v\textsubscript{GBP}_t, t) dt + s\textsuperscript{v}(v\textsubscript{GBP}_t, t) dW\textsuperscript{v\_GBP}_t + \omega\textsuperscript{v}(v\textsubscript{GBP}_t, t) dJ\textsuperscript{v\_GBP}_t
\end{equation}

\paragraph{SC Choice}

\begin{equation}
d\rho_t = m\textsuperscript{$\rho$}(\rho_t, t) dt + s\textsuperscript{$\rho$}(\rho_t, t) dW\textsuperscript{$\rho$}_t + \omega\textsuperscript{$\rho$}(\rho_t, t) dJ\textsuperscript{$\rho$}_t
\end{equation}

\paragraph{SER Choice}

\begin{equation}
dFX\textsubscript{GBP}(t) = m\textsuperscript{f}(FX\textsubscript{GBP}(t), t) dt + s\textsuperscript{f}(FX\textsubscript{GBP}(t), t) dW\textsuperscript{f}_t + \omega\textsuperscript{f}(FX\textsubscript{GBP}(t), t) dJ\textsuperscript{f}_t
\end{equation}

\subsubsection{Correlation between Brownian Motions}
\begin{equation}
\begin{aligned}
dW_{USD}(t) &= \rho(t) dW_{GBP}(t) + \sqrt{1 - \rho(t)^2} dZ(t) \\
dW_{FX}(t) &= dZ_{FX}(t)
\end{aligned}
\end{equation}
Here, \( dZ(t) \) and \( dZ_{FX}(t) \) are independent standard Brownian motions.

\subsubsection{Payoff}
\begin{equation}
\text{Payoff} = \max \left( S_{USD}(T) - K_1 , S_{GBP}(T) \cdot FX(T) - K_2 , 0 \right)
\end{equation}
\\

\noindent Here, \( dJ\textsuperscript{v}_t \) denotes the jump component of volatility, \( dJ\textsuperscript{$\rho$}_t \) denotes the jump component of correlation, and \( dJ\textsuperscript{f}_t \) denotes the jump component for the FX rate, as defined in Section 2. Our choice of the SV, SC, and SER models determines the m, s, and $\omega$ functions above. For example, GARCH-Jump SV has $m\textsuperscript{v}(v_t, t)$ = $\beta_0 + (\beta_1 - 1) v_t$, $s\textsuperscript{v}(v_t, t)$ = $0$, and $\omega\textsuperscript{v}(v_t, t)$ = $\beta_2 v_t $. Similarly, we can write the general equations for our model for Case 2 as shown below. 

\subsection{Case 2: Three Underlying Assets - Two FX Rates}
\subsubsection{Variables}
\begin{itemize}
    \item \( S_{GBP}(t) \): Price of the GBP asset.
    \item \( S_{USD}(t) \): Price of the USD asset.
    \item \( S_{EUR}(t) \): Price of the EUR asset.
    \item \( FX_{GBP}(t) \): FX rate for USD/GBP.
    \item \( FX_{EUR}(t) \): FX rate for USD/EUR.
\end{itemize}

\subsubsection{Stochastic Differential Equations}

\paragraph{Underlying Asset Prices}
\begin{equation}
dS_{USD}(t) = S_{USD}(t) \left( r_d dt + \sqrt{v_{USD}(t)} dW_{USD}(t) \right)
\end{equation}

\begin{equation}
dS_{GBP}(t) = S_{GBP}(t) \left( r_{f_1} dt + \sqrt{v_{GBP}(t)} dW_{GBP}(t) \right)
\end{equation}

\begin{equation}
dS_{EUR}(t) = S_{EUR}(t) \left( r_{f_2} dt + \sqrt{v_{EUR}(t)} dW_{EUR}(t) \right)
\end{equation}

\paragraph{SV Choice (same process for all underlying assets)}

\begin{equation}
dv\textsubscript{USD}_t = m\textsuperscript{v}(v\textsubscript{USD}_t, t) dt + s\textsuperscript{v}(v\textsubscript{USD}_t, t) dW\textsuperscript{v\_USD}_t + \omega\textsuperscript{v}(v\textsubscript{USD}_t, t) dJ\textsuperscript{v\_USD}_t
\end{equation}

\begin{equation}
dv\textsubscript{GBP}_t = m\textsuperscript{v}(v\textsubscript{GBP}_t, t) dt + s\textsuperscript{v}(v\textsubscript{GBP}_t, t) dW\textsuperscript{v\_GBP}_t + \omega\textsuperscript{v}(v\textsubscript{GBP}_t, t) dJ\textsuperscript{v\_GBP}_t
\end{equation}

\begin{equation}
dv\textsubscript{EUR}_t = m\textsuperscript{v}(v\textsubscript{EUR}_t, t) dt + s\textsuperscript{v}(v\textsubscript{EUR}_t, t) dW\textsuperscript{v\_EUR}_t + \omega\textsuperscript{v}(v\textsubscript{EUR}_t, t) dJ\textsuperscript{v\_EUR}_t
\end{equation}

\paragraph{SC Choice (between both pairs, USD-GBP and USD-EUR)}

\begin{equation}
d\rho_t = m\textsuperscript{$\rho$}(\rho_t, t) dt + s\textsuperscript{$\rho$}(\rho_t, t) dW\textsuperscript{$\rho$}_t + \omega\textsuperscript{$\rho$}(\rho_t, t) dJ\textsuperscript{$\rho$}_t
\end{equation}

\paragraph{SER Choice (same process for both FX Rates, USD/GBP, \& USD/EUR)}
\begin{equation}
dFX\textsubscript{GBP}(t) = m\textsuperscript{f}(FX\textsubscript{GBP}(t), t) dt + s\textsuperscript{f}(FX\textsubscript{GBP}(t), t) dW\textsuperscript{f1}_t + \omega\textsuperscript{f}(FX\textsubscript{GBP}(t), t) dJ\textsuperscript{f1}_t
\end{equation}

\begin{equation}
dFX\textsubscript{EUR}(t) = m\textsuperscript{f}(FX\textsubscript{EUR}(t), t) dt + s\textsuperscript{f}(FX\textsubscript{EUR}(t), t) dW\textsuperscript{f2}_t + \omega\textsuperscript{f}(FX\textsubscript{EUR}(t), t) dJ\textsuperscript{f2}_t
\end{equation}

\subsubsection{Correlation between Brownian Motions}
\begin{equation}
\begin{aligned}
dW_{USD}(t) &= \rho(t) dW_{GBP}(t) + \sqrt{1 - \rho(t)^2} dZ(t) \\
dW_{EUR}(t) &= \rho(t) dW_{GBP}(t) + \sqrt{1 - \rho(t)^2} dZ_{EUR}(t) \\
dW_{FX_{USD/GBP}}(t) &= dZ_{FX_{USD/GBP}}(t) \\
dW_{FX_{USD/EUR}}(t) &= dZ_{FX_{USD/EUR}}(t)
\end{aligned}
\end{equation}
Here, \( dZ(t) \), \( dZ_{EUR}(t) \), \( dZ_{FX_{USD/GBP}}(t) \), and \( dZ_{FX_{USD/EUR}}(t) \) are independent standard Brownian motions.

\subsubsection{Payoff}
\begin{equation}
\begin{aligned}
\text{Payoff} = \max \left( S_{USD}(T) - K_1, S_{GBP}(T) \cdot FX_{GBP}(T) - K_2, \right. \\
\left. S_{EUR}(T) \cdot FX_{EUR}(T) - K_3, 0 \right)
\end{aligned}
\end{equation}
\\

\noindent Hence, as highlighted before, we correlate the Brownian motions of the processes of the underlying assets and model this correlation as stochastic.  Also, we model the volatility of the underlying as stochastic and independent from the level of the underlying. Moreover, the FX rates are modeled as stochastic and independent from each other as well as from the level of their respective underlying asset. In the simulation, we use different sets of increments of Brownian motion for all the stochastic processes and only introduce a correlation between those of the underlying assets. This correlation, between the Brownian motions of USD and GBP assets as well as between the Brownian motions of the USD and EUR assets, is the same in our experiments. In all of the SC models, we tested $\omega\textsuperscript{$\rho$}(\rho_t, t) $ = 0 as we do not investigate the presence of jumps in correlation. 

\section{DISCRETIZATION of SDEs \& MC SIMULATION}
We test all 60 (= 5*4*3) combinations of different choices of the SV, SC, and SER models with three different discretization schemes and we use the Euler-Maruyama scheme only for the SDEs of the underlying assets throughout. The Euler-Maruyama Scheme is the simplest computationally as it only includes the first three terms of the Ito-Taylor expansion applied to the SDE. However, it is expected to yield limited accuracy since this approximation expands the drift term to \(O(\Delta t)\) but only expands the diffusion term to \(O(\sqrt{\Delta t})\). The Milstein scheme should yield improved accuracy since a second diffusion term is added, expanding the diffusion term to \(O(\Delta t)\) as well. Although the Milstein scheme has a higher order, its main drawback is that we need to compute the first derivative of the volatility function. This may not always be possible, or it may be computationally expensive. The Runge-Kutta scheme can be used to alleviate this issue while maintaining this higher-order by leveraging the Runge-Kutta approximation of the derivative required in the Milstein scheme. Higher-order Runge-Kutta schemes can be derived by including a more detailed approximation of this derivative. The SDE of a general Itô diffusion \({I_t}\) is \( dI_t = a(I_t,t)dt + b(I_t,t)dW_t \) and can be discretized via the three alternatives as described below. Here, when discretizing and simulating the increments of Brownian motion, \(\Delta W_j = W_{t_{j+1}} - W_{t_j}\) \(\stackrel{i.i.d}{\sim}\) \(N(0, \Delta t)\), we write \(\Delta W_j\) = \(\sqrt{\Delta t} Z_j\) with \( Z_j \stackrel{i.i.d}{\sim} N(0, 1)\), \(\forall j\), and \( \Delta t = t_{j+1} - t_{j} \). To apply the Euler-Maruyama, Milstein, and Runge-Kutta discretization schemes for SDEs that include an additional jump term, such as those in the GARCH-Jump, Bates Stochastic Volatility (SV), and Exponential Lévy FX rate models, these schemes need to be adjusted as described below. A Itô process with an additional jump component can be written as:

\begin{equation}
    dI_t = a(I_t, t)dt + b(I_t, t)dW_t + c(I_t, t)dJ_t
\end{equation}

\noindent where \(dJ_t\) represents the jump term, often modeled as a compound Poisson process. Indeed, in our paper, the jumps are modeled as compound Poisson processes as discussed in section 2.1.3. Here, we have \(\Delta J_j = J_{t_{j+1}} - J_{t_j}\), where \(\Delta J_j\) represents the cumulative effect of jumps in the interval \([t_j, t_{j+1}]\). Specifically, \(\Delta J_j = \sum_{k=1}^{N_j} Y_k\), where \(N_j \sim \text{Poisson}(\lambda \Delta t)\) represents the number of jumps occurring in the time interval \(\Delta t = t_{j+1} - t_j\), and each \(Y_k \stackrel{i.i.d}{\sim} N(\mu_J, \sigma_J^2)\) represents the jump sizes drawn from a normal distribution with mean \(\mu_J\) and variance \(\sigma_J^2\). Thus, \(\Delta J_j\) \(\sim \text{Poisson}(\lambda \cdot \Delta t) \times \mathcal{N}(\mu_J, \sigma_J^2) \).

\subsection{Euler-Maruyama Scheme}
\begin{equation}
\hat{I}_{t_{j+1}} = \hat{I}_{t_j} + a(\hat{I}_{t_j}, t_j) \Delta t + b(\hat{I}_{t_j}, t_j) \Delta W_j
\end{equation}
\begin{equation}
= \hat{I}_{t_j} + a(\hat{I}_{t_j}, t_j) \Delta t + b(\hat{I}_{t_j}, t_j) \sqrt{\Delta t} Z_j
\end{equation}

\subsection{Euler-Maruyama Scheme with Jumps}
\begin{equation}
    \hat{I}_{t_{j+1}} = \hat{I}_{t_j} + a(\hat{I}_{t_j}, t_j) \Delta t + b(\hat{I}_{t_j}, t_j) \sqrt{\Delta t} Z_j + c(\hat{I}_{t_j}, t_j) \Delta J_j
\end{equation}

\subsection{Milstein Scheme}
\begin{equation}
\hat{I}_{t_{j+1}} = \hat{I}_{t_j} + a(\hat{I}_{t_j}, t_j) \Delta t + b(\hat{I}_{t_j}, t_j) \Delta W_j + \frac{1}{2} b(\hat{I}_{t_j}, t_j) b'(\hat{I}_{t_j}, t_j) \left[ (\Delta W_j)^2 - \Delta t \right]
\end{equation}
\begin{equation}
= \hat{I}_{t_j} + a(\hat{I}_{t_j}, t_j) \Delta t + b(\hat{I}_{t_j}, t_j) \sqrt{\Delta t} Z_j + \frac{1}{2} b(\hat{I}_{t_j}, t_j) b'(\hat{I}_{t_j}, t_j) \Delta t (Z_j^2 - 1)
\end{equation}

\subsection{Milstein Scheme with Jumps}
\begin{equation}
    \hat{I}_{t_{j+1}} = \hat{I}_{t_j} + a(\hat{I}_{t_j}, t_j) \Delta t + b(\hat{I}_{t_j}, t_j) \Delta W_j + \frac{1}{2} b(\hat{I}_{t_j}, t_j) b'(\hat{I}_{t_j}, t_j) \Delta t (Z_j^2 - 1) + c(\hat{I}_{t_j}, t_j) \Delta J_j
\end{equation}

\subsection{Runge-Kutta Scheme}
\begin{equation}
\hat{I}_{t_{j+1}} = \hat{I}_{t_j} + a(\hat{I}_{t_j}, t_j) \Delta t + b(\hat{I}_{t_j}, t_j) \Delta W_j + 
\end{equation}
\begin{equation*}
    \frac{1}{2 \sqrt{\Delta t}} \left[ b(\tilde{I}_{t_j}, t_j) - b(\hat{I}_{t_j}, t_j) \right] \left[ (\Delta W_j)^2 - \Delta t \right]  
\end{equation*}
\begin{equation}
    = \hat{I}_{t_j} + a(\hat{I}_{t_j}, t_j) \Delta t + b(\hat{I}_{t_j}, t_j) \sqrt{\Delta t} Z_j + 
\end{equation}
\begin{equation*}
    \frac{1}{2} \left[ b(\tilde{I}_{t_j}, t_j) - b(\hat{I}_{t_j}, t_j) \right] \sqrt{\Delta t} \left[ (\Delta Z_j)^2 - 1 \right]
\end{equation*}
where
\begin{equation}
\tilde{I}_{t_j} = \hat{I}_{t_j} + a(\hat{I}_{t_j}, t_j) \Delta t + b(\hat{I}_{t_j}, t_j) \sqrt{\Delta t}
\end{equation}

\subsection{Runge-Kutta Scheme with Jumps}
\begin{equation}
    \tilde{I}_{t_j} = \hat{I}_{t_j} + a(\hat{I}_{t_j}, t_j) \Delta t + b(\hat{I}_{t_j}, t_j) \sqrt{\Delta t} + c(\hat{I}_{t_j}, t_j) \Delta J_j
\end{equation}

\begin{equation}
    \hat{I}_{t_{j+1}} = \hat{I}_{t_j} + a(\hat{I}_{t_j}, t_j) \Delta t + b(\hat{I}_{t_j}, t_j) \sqrt{\Delta t} Z_j + 
\end{equation}
\begin{equation*}
    \frac{1}{2} \left[ b(\tilde{I}_{t_j}, t_j) - b(\hat{I}_{t_j}, t_j) \right] \sqrt{\Delta t} \left[ (\Delta Z_j)^2 - 1 \right] + c(\hat{I}_{t_j}, t_j) \Delta J_j
\end{equation*}

\noindent In summary, to handle an additional jump term in our SDEs we add the jump component directly to the discretization for the Euler-Maruyama and Milstein schemes, and the Runge-Kutta scheme also requires calculating the intermediate values including the jump component, and then adjusting the final update accordingly. These modifications ensure that each discretization scheme accurately captures the effects of both continuous and jump components in the SDEs.  

\subsection{Simulation Procedure}
\begin{enumerate}
    \item \textbf{Initialize paths:}
    Set initial values for \( S_{GBP}(0) \), \( S_{USD}(0) \), \( FX(0) \), \( \rho(0) \), \( v_{GBP}(0) \), and \( v_{USD}(0) \). Set T (time to expiration in years and \(n_{steps}\) per path in days is T*252), N (number of paths in simulation), set increment dt = T/\(n_{steps}\) for daily time series. 

    \item \textbf{Simulate paths over time:}
    \begin{itemize}
        \item For each time step, generate Brownian increments and jump instances.
        \item Update the volatility paths using the specified volatility model and discretization scheme.
        \item Update the correlation paths using the specified correlation model and discretization scheme.
        \item Update the FX rate paths using the specified model and discretization scheme.
        \item Update the underlying asset paths using the simulated volatility and correlation paths according to the Euler-Maruyama discretization scheme. 
    \end{itemize}

    \item \textbf{Calculate payoffs:}
    Use the final asset prices and FX rate values to calculate the payoff for each simulated path. 
    For Case 1, the payoff for each path is calculated as:
    \begin{equation}
    \text{Payoff}_i = \max \left( S_{USD,i}(T) - K_1 , S_{GBP,i}(T) \cdot FX_i(T) - K_2 , 0 \right)
    \end{equation}

    For Case 2, the payoff for each path is calculated as:
    \begin{equation}
    \begin{aligned}
        \text{Payoff}_i = \max \left( S_{USD,i}(T) - K_1, S_{GBP,i}(T) \cdot FX_{USD/GBP,i}(T) - K_2, \right. \\
        \left. S_{EUR,i}(T) \cdot FX_{USD/EUR,i}(T) - K_3, 0 \right) 
    \end{aligned}
    \end{equation}

    \item \textbf{Discount and average payoffs:}
    Discount the payoffs to present value and average them to get the option price. The discounted option price is given by:
    \begin{equation}
    \text{Option Price} = e^{-r_d T} \cdot \frac{1}{N} \sum_{i=1}^{N} \text{Payoff}_i
    \end{equation}
    where \( N \) is the number of simulated paths, \( r_d \) is the domestic risk-free rate, and \( T \) is the time to maturity.
\end{enumerate}

\subsection{Variance Reduction - Antithetic Variates}
Variance reduction methods seek to improve the efficiency of an estimator, enabling more accurate results for a given number of simulations, N. Some choices of variance reduction that could be applied to our Monte Carlo simulations include control variates, which leverage the known expectation of auxiliary variables to reduce variance, and randomized Quasi-Monte Carlo (QMC), which replaces purely random sampling with low-discrepancy sequences (e.g. Halton, Sobol, or Fauer) to enhance convergence properties. In this paper, we implement antithetic variates as a form of variance reduction to improve our Monte Carlo simulations. Antithetic variates aim to reduce the variance of the estimator by introducing a negative correlation between paired simulations, thereby improving the convergence rate of the Monte Carlo estimator without increasing the number of simulations. In our context of Monte Carlo simulations involving stochastic differential equations (SDEs), this involves generating antithetic paths by using the negative of the original Brownian motion increments. Then, by averaging the results from the original and antithetic paths, the variance of the estimator is reduced. Specifically, let \( X \) be an estimator of the option price based on the original simulations, and \( X' \) be the estimator based on the antithetic simulations. The combined estimator is then:

\begin{equation} 
\hat{C} = \frac{1}{2} (X + X')
\end{equation}

\noindent Now, since \( X \) and \( X' \) are negatively correlated, the variance of \( \hat{C} \) is reduced compared to the variance of either \( X \) or \( X' \) alone. The variance of our estimator can be expressed as follows: 

\begin{equation} 
\text{Var}(\hat{C}) = \frac{1}{2} \left( \text{Var}(X + X') \right) = \frac{1}{4} \left( \text{Var}(X) + \text{Var}(X') + 2 \cdot \text{Cov}(X, X') \right)
\end{equation}

\noindent and our X's both have equal variance so this expression can be simplified further as follows: 
\begin{equation} 
\text{Var}(\hat{C}) = \frac{1}{4} \left( (2*2) \text{Var}(X) + 2 \cdot \text{Cov}(X, X') \right) = \text{Var}(X) + \frac{1}{2} \text{Cov}(X, X')
\end{equation}

\noindent which means that the variance of our antithetic estimator is less than that of the original estimator by \( | \frac{1}{2} Cov(X, X') | \) if the covariance between the antithetic and original versions of the random variables is negative: 
\begin{equation}
    \text{Var}(\hat{C}) < \text{Var}(X) \iff Cov(X, X') < 0 
\end{equation}

\noindent Hence, the effectiveness of the antithetic variates method depends on the extent of the negative correlation between the estimators \( X \) and \( X' \). The more negatively correlated they are, the greater the variance reduction. Since \( \text{Var}(X) = \text{Var}(X') \), we can rewrite this reduction in variance as follows: 
\begin{equation}
    | \frac{1}{2} \text{Cov}(X, X') | =  | \frac{1}{2} \rho \sigma_{x} \sigma_{x'} | = | \frac{1}{2} \rho \text{Var}(X) |
\end{equation}

\subsubsection{Antithetic Brownian Increments}

To incorporate antithetic variates into our simulation process, for each path in the simulation, we generate a corresponding antithetic path by using the negative of the samples of a standard normal random variable in the Brownian increments used in the original path. This approach is applied to all the Brownian motions involved in the simulation. Specifically, if we let \( \Delta W_j \) be the Brownian increments generated for the original path at time step \( t_j \). Then, by writing  \(Z'_j\) = \( - Z_j\), we get that the antithetic increments \( \Delta W'_j \) are given by:

\begin{equation*}
    \Delta W'_j = \sqrt{\Delta t} Z'_j = - \sqrt{\Delta t} Z_j = -\Delta W_j
\end{equation*} 

\subsubsection{Modification of the Simulation Procedure}
We modify the steps of our simulation process outlined in section 4.7 as described below to incorporate antithetic variates. Step 1 is the same as in the original version, but steps 2 and 3 change slightly as we have to include the generation of the antithetic versions for all Brownian increments and updating all paths (including the antithetic versions), respectively. Step 4 also mainly stays the same but we need to compute the payoffs for both the antithetic paths and the normal paths. Then, the key adjustment is in changing step 5 as follows: 

\noindent\textbf{New step 5 - discount and average \textit{combined} payoffs:}
Compute the combined payoff for each pair of original and antithetic sets of paths as:
\begin{equation}
    \overline{\text{Payoff}}_i = \frac{1}{2} \left( \text{Payoff}_i + \text{Payoff}'_i \right)
\end{equation}
As before, discount the combined payoffs and average them to obtain the option price estimate:
\begin{equation}
    \text{Option Price} = e^{-r_d T} \cdot \frac{1}{N} \sum_{i=1}^{N} \overline{\text{Payoff}}_i
\end{equation}

\noindent In both case 1 and case 2 of our options, since our payoff functions are monotonic in the asset prices (\& FX rates) and all the SDEs we use are linear in the Brownian increments, the negative correlation introduced at the level of the Brownian increments propagates through to the terminal asset prices, and the option payoffs. Also, the symmetry of the Gaussian distribution guarantees that when we negate a (zero-mean) Normal random variable it is still distributed according to the same distribution. Because of this, for any of our SDEs, the solution of the version modified with antithetic Brownian increments has the same distribution as the original SDE's solution. Indeed, with this, it can be shown that the antithetic and original option price estimates are identically distributed and have a negative covariance \cite{Genz2017} - as required for variance reduction via antithetic variates. 

\section{RESULTS COMPARISON \& DISCUSSION}
We use the number of simulation paths with N = 500,000 for all of the simulations in this study. This number of realizations performed well in the preliminary testing of the code. The lists of parameter values used for the simulations of 2021 and 2022 starting dates, respectively, are shown in the last two figures. To compare the performance of the discretization schemes and model variants, we compute the standard deviation of the option value estimates and 95\% confidence intervals (using z* = 1.96). Then, all 180 model variants are ordered from lowest standard deviation to highest, and we compute the average price estimate of the 40 best variants to be the 'true' value of the option. Then, all the models are re-ordered by the lowest to highest percentage error from this value, and the top 30 are displayed in bold in the results tables. Figures 2 and 3 below display the top 30 models by percentage error (PE) for cases 1 and 2 respectively and for both start dates we test (2021 on the left and 2022 on the right subplots). 
\begin{figure}[h]
    \centering
    \includegraphics[width=\linewidth]{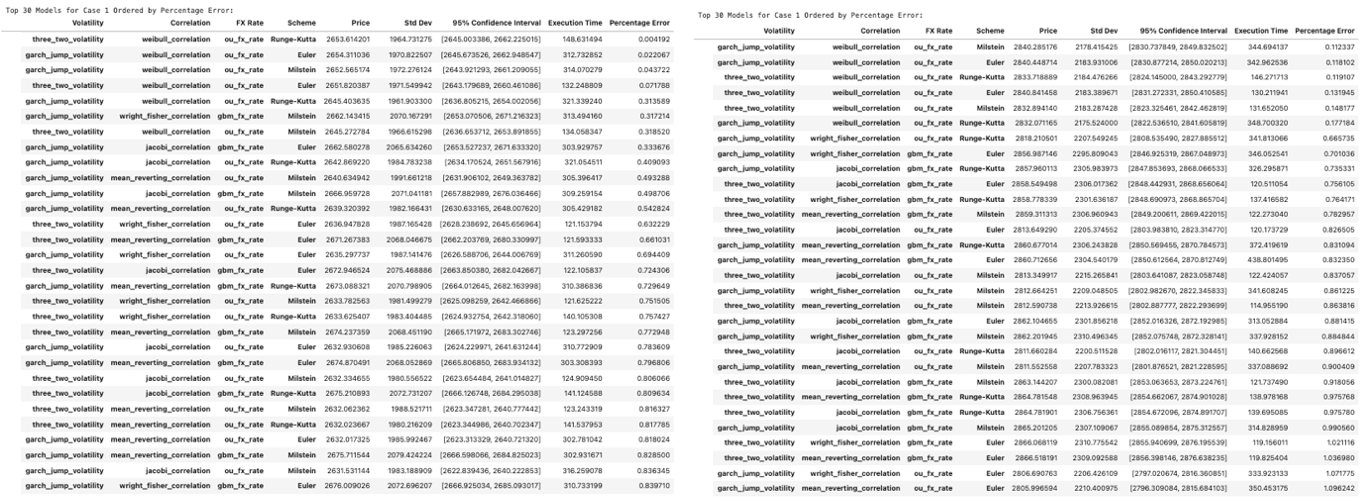} % 
    \caption{Case 1 top 30 models by percentage error for 2021 (left) and 2022 (right) start dates.}
    \label{fig:Case_1_top_30_by_PE}
\end{figure}

\begin{figure}[h]
    \centering
    \includegraphics[width=\linewidth]{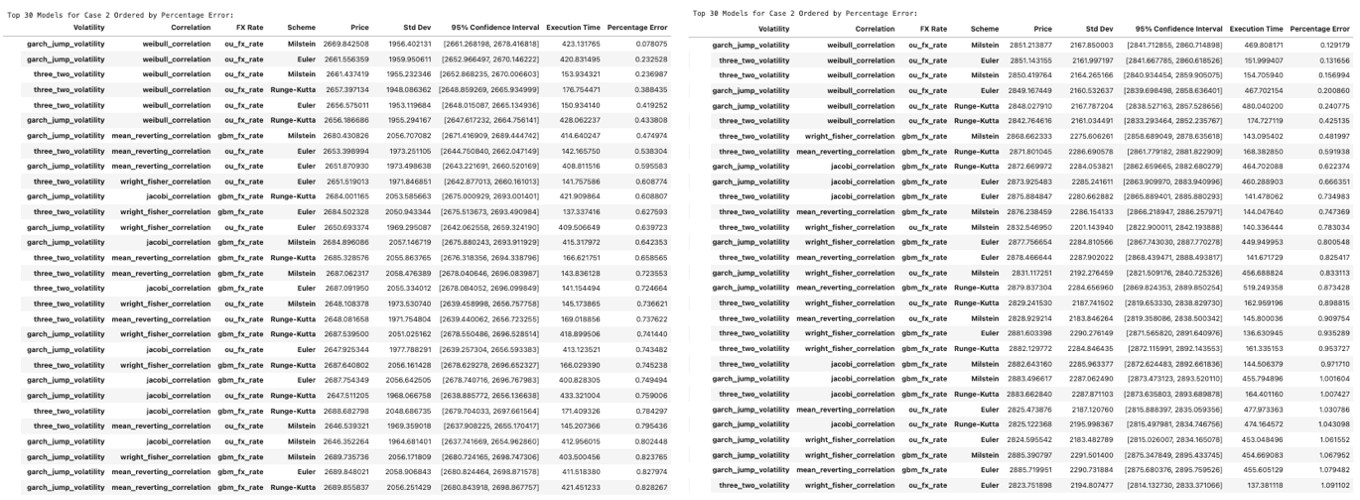} % 
    \caption{Case 2 top 30 models by percentage error for 2021 (left) and 2022 (right) start dates.}
    \label{fig:Case_2_top_30_by_PE}
\end{figure}

Figures 4-19 (inclusive) show the outputs for all model variants and discretization schemes ordered by PE. From our simulations, the target values (average of 40 best models by standard deviation) for case 1 of the option with 2021 and 2022 start dates were \$ 2653.73 and \$ 2837.10, respectively (2 d.p.). For case 2 of the option, our simulations yield the values \$ 2667.76 and \$ 2854.90 for 2021 and 2022 start dates, respectively (2 d.p.). It makes sense that these options were slightly more expensive with the 2022 starting date since all the assets were significantly more expensive on 2022-01-04 than on 2021-01-4, as shown in Figure 1. These are call options, so the strike is subtracted from the respective underlying (weighted by exchange rate) in each component of the payoff. This means the option price should be greater if the prices of the underlying assets are all higher ceteris paribus. Comparing models by percentage error relative to this calculated value is an attempt to perform model selection even though we do not have real observed prices, another pricing method, or a closed-form/series solution to compare our MC simulation prices and performance to. As shown in Figures 2 and 3, the best-performing combination of SV, SC, and SER models for case 1 is (GARCH-Jump, Weibull, OU) for both start dates of the option. For case 2, the best-performing combination is also (GARCH-Jump, Weibull, OU). The 3/2 SV model was a close runner-up in our simulations but a higher number of the top 10 models across both cases and start dates of the options include GARCH-Jump SV. The combination of the Weibull SC and OU SER models performs better than all the alternatives we tested for SC and SER choices. Figures 33-41 show plots of some sample paths from each of the (9) SDEs in our best model over 252 time steps. 
\\

As a benchmark, we also test the performance of all model variants with constant correlation. Figures 20-23 (inclusive) show the outputs of the MC simulations with constant correlation with the tables of the model variants ordered by percentage error. The best-performing model variants with constant correlation achieve 0.436 (case 1 with 2021 start), 0.154 (case 2 with 2021 start), 0.147 (case 1 with 2022 start), and 0.431 (case 2 with 2022 start) percentage error. These best-performing model variants have the following combinations of SV, SER, and discretization scheme: (3/2, GBM, Runge-Kutta) for case 1 with 2021 start, (3/2, GBM, Euler) for case 2 with 2021 start, (GARCH-Jump, GBM, Runge-Kutta) for case 1 with 2022 start, and (GARCH-Jump, GBM, Milstein) for case 2 with 2022 start. For case 1 of the option with a 2021 start date, this performance places the best constant correlation model 13th overall by percentage error when compared to the results of all models with stochastic correlation. This indicates that the inclusion of stochastic correlation significantly improves pricing via MC simulation. However, for case 2 with the same start date and both cases of the option with the 2022 start date, the best performing constant correlation models place 4th, 4th, and 5th, respectively, when compared to the 30 best model variants with SC. As highlighted by Ma, market data analysis reveals that implied correlation often deviates from realized correlation, indicating a non-zero correlation risk premium \cite{Ma2009} (Buraschi, Porchia, and Trojani, 2006 \cite{Buraschi2006}; Driessen, Maenhout, and Vilkov, 2006 \cite{Driessen2006}). Moreover, Figure 20 shows that correlations seem to fluctuate randomly over time and are far from constant. This evidence supports the inclusion of random correlation structures in derivative pricing models.
\\

The execution times of the simulations of all of the model variants are also recorded in the code and displayed in seconds. The Euler scheme generally exhibits faster execution times and performs well. The Milstein and Runge-Kutta schemes show lower stdevs, indicating higher accuracy compared to the Euler scheme. The Runge-Kutta scheme often shows the lowest stdevs, indicating the best accuracy, albeit with longer execution times. Based on our simulations, the Milstein discretization scheme offers the best balance between execution times and lower standard deviations. However, the Runge-Kutta scheme should be used when the derivative of the coefficient of the diffusion term is not available or hard to compute. In addition, the simulations for the 2022-2023 period generally showed slightly increased execution times and wider confidence intervals compared to the 2021-2022 period, reflecting the increased volatility and market uncertainty during that time frame. This is also reflected in the increased rolling correlations. Also, the simulations took slightly longer on average for case 2 of the option than for case 1 as it involves more processes to be simulated. For the same reason, price estimates for case 2 of the option generally have wider CIs on average than those for case 1. 
\\

\section{HEDGING CORRELATION RISKS}
In the context of our foreign equity quanto call options, Cora and Gora are metrics used to quantify and manage the correlation risks associated with the inclusion of multiple correlated underlying assets in the payoffs of these complex derivatives. Cora (correlation delta) measures the sensitivity of the option's value to changes in the correlation between the underlying assets and Gora (correlation gamma) measures the second-order sensitivity. It is a crucial parameter for understanding how variations in correlation impact the option's price, especially when the correlation itself is stochastic. By using Cora and Gora, traders and risk managers can better understand and manage the impact of correlation changes on the option prices, ensuring more effective hedging strategies for portfolios that include basket foreign equity quanto call options such as those explored in our study \cite{Karlsson2009}. Cora is defined as the first-order partial derivative of the option price (discounted expectation of payoff) with respect to correlation: 

\begin{equation}
\text{Cora} = \frac{\partial C}{\partial \rho}
\end{equation}

\noindent Gora is defined as the second-order partial derivative of the option price with respect to correlation: 
\begin{equation}
\text{Gora} = \frac{\partial^2 C}{\partial \rho^2}
\end{equation}

\noindent The discount factor can be taken outside of these derivatives, so the main task of deriving Cora and Gora becomes taking derivatives of the option's payoff and applying the chain rule. Also, in our models the SC only applies to the BMs driving the stochastic processes of the underlying assets and the BMs driving the FX rate processes are independent of this stochastic correlation process. This allows us to simplify the formulas for Cora and Gora as all partial derivatives of the FX rate processes with respect to this SC are 0. 

\subsection{Case 1: Single FX Rate}
For the single FX rate case, where the option payoff is given by:
\begin{equation}
\text{Payoff} = \max \left( S_{USD}(T) - K_1 , S_{GBP}(T) \cdot FX(T) - K_2 , 0 \right)
\end{equation}

\noindent The Cora and Gora metrics can be derived as follows:
Cora measures how sensitive the option's value is to changes in the correlation between the GBP asset price and the USD asset price. Mathematically, this sensitivity is expressed as:
\begin{equation}
\text{Cora}{\text{}} = \frac{\partial C}{\partial \rho_{GBP,USD}}
\end{equation}

\noindent Gora measures the rate of change of the sensitivity (Cora) with respect to the correlation between the GBP and USD asset prices. It captures the curvature of the option price with respect to correlation changes, indicating how Cora itself responds as the correlation changes:
\begin{equation}
\text{Gora}{\text{}} = \frac{\partial^2 C}{\partial \rho_{GBP,USD}^2}
\end{equation}

\noindent Using the chain rule, the first partial derivative of \(C\) with respect to \(\rho_{GBP,USD}\) is:
\begin{equation}
\frac{\partial C}{\partial \rho_{GBP,USD}} = \frac{\partial C}{\partial S_{GBP}} \cdot \frac{\partial S_{GBP}}{\partial \rho_{GBP,USD}} + \frac{\partial C}{\partial S_{USD}} \cdot \frac{\partial S_{USD}}{\partial \rho_{GBP,USD}} + \frac{\partial C}{\partial FX} \cdot \frac{\partial FX}{\partial \rho_{GBP,USD}}
\end{equation}

\noindent then, since the partial derivative of the FX rate process with respect to the correlation is 0, we get: 
\begin{align}
\text{Cora}_{\text{}} &= \frac{\partial C}{\partial S_{GBP}} \cdot \frac{\partial S_{GBP}}{\partial \rho_{GBP,USD}} + \frac{\partial C}{\partial S_{USD}} \cdot \frac{\partial S_{USD}}{\partial \rho_{GBP,USD}}
\end{align}

\noindent For the second partial derivative, we get:
\begin{align}
\frac{\partial^2 C}{\partial \rho_{GBP,USD}^2}  
&= \frac{\partial}{\partial \rho_{GBP,USD}} \left( \frac{\partial C}{\partial S_{GBP}} \cdot \frac{\partial S_{GBP}}{\partial \rho_{GBP,USD}} + \frac{\partial C}{\partial S_{USD}} \cdot \frac{\partial S_{USD}}{\partial \rho_{GBP,USD}} \right) \nonumber\\ 
&= \frac{\partial}{\partial \rho_{GBP,USD}} \left( \frac{\partial C}{\partial S_{GBP}} \right) \cdot \frac{\partial S_{GBP}}{\partial \rho_{GBP,USD}} + \frac{\partial C}{\partial S_{GBP}} \cdot \frac{\partial}{\partial \rho_{GBP,USD}} \left( \frac{\partial S_{GBP}}{\partial \rho_{GBP,USD}} \right) \nonumber\\ 
&+ \frac{\partial}{\partial \rho_{GBP,USD}} \left( \frac{\partial C}{\partial S_{USD}} \right) \cdot \frac{\partial S_{USD}}{\partial \rho_{GBP,USD}} + \frac{\partial C}{\partial S_{USD}} \cdot \frac{\partial}{\partial \rho_{GBP,USD}} \left( \frac{\partial S_{USD}}{\partial \rho_{GBP,USD}} \right) 
\end{align}

\noindent Simplifying fully, we get:
\begin{align}
\text{Gora}_{\text{}} &= \frac{\partial^2 C}{\partial S_{GBP} \partial \rho_{GBP,USD}} \cdot \frac{\partial S_{GBP}}{\partial \rho_{GBP,USD}} + \frac{\partial C}{\partial S_{GBP}} \cdot \frac{\partial^2 S_{GBP}}{\partial \rho_{GBP,USD}^2} \nonumber\\
&+ \frac{\partial^2 C}{\partial S_{USD} \partial \rho_{GBP,USD}} \cdot \frac{\partial S_{USD}}{\partial \rho_{GBP,USD}} + \frac{\partial C}{\partial S_{USD}} \cdot \frac{\partial^2 S_{USD}}{\partial \rho_{GBP,USD}^2} 
\end{align}

\subsection{Case 2: Two FX Rates}
For the two FX rates case, the option payoff is given by:

\begin{equation}
\text{Payoff} = \max \left( S_{USD}(T) - K_1, S_{GBP}(T) \cdot FX_{USD/GBP}(T) - K_2,  S_{EUR}(T) \cdot FX_{USD/EUR}(T) - K_3, 0 \right)
\end{equation}

\noindent Here, Cora and Gora metrics need to account for multiple correlations: between GBP and USD asset prices and between EUR and USD asset prices. The Cora for each of these correlations would be defined as:
\begin{equation}
\text{Cora}_{\text{GBP,USD}} = \frac{\partial C}{\partial \rho_{GBP,USD}}
\end{equation}

\begin{equation}
\text{Cora}_{\text{EUR,USD}} = \frac{\partial C}{\partial \rho_{EUR,USD}}
\end{equation}

\noindent Similarly, the Gora metrics would measure the second-order sensitivity for each of these correlations, indicating how each Cora changes with respect to changes in the corresponding correlations:
\begin{equation}
\text{Gora}_{\text{GBP,USD}} = \frac{\partial^2 C}{\partial \rho_{GBP,USD}^2}
\end{equation}

\begin{equation}
\text{Gora}_{\text{EUR,USD}} = \frac{\partial^2 C}{\partial \rho_{EUR,USD}^2}
\end{equation}

\noindent Using the chain rule, the first partial derivative of \(C\) with respect to \(\rho_{GBP,USD}\) is:
\begin{equation}
\begin{aligned}
\frac{\partial C}{\partial \rho_{GBP,USD}} &= \frac{\partial C}{\partial S_{GBP}} \cdot \frac{\partial S_{GBP}}{\partial \rho_{GBP,USD}} + \frac{\partial C}{\partial S_{USD}} \cdot \frac{\partial S_{USD}}{\partial \rho_{GBP,USD}} \\
&+ \frac{\partial C}{\partial S_{EUR}} \cdot \frac{\partial S_{EUR}}{\partial \rho_{GBP,USD}} + \frac{\partial C}{\partial FX_{USD/GBP}} \cdot \frac{\partial FX_{USD/GBP}}{\partial \rho_{GBP,USD}}
\end{aligned}
\end{equation}

\noindent Similarly, the first partial derivative of \(C\) with respect to \(\rho_{EUR,USD}\) is:
\begin{equation}
\begin{aligned}
\frac{\partial C}{\partial \rho_{EUR,USD}} &= \frac{\partial C}{\partial S_{EUR}} \cdot \frac{\partial S_{EUR}}{\partial \rho_{EUR,USD}} + \frac{\partial C}{\partial S_{USD}} \cdot \frac{\partial S_{USD}}{\partial \rho_{EUR,USD}} \\
&+ \frac{\partial C}{\partial S_{GBP}} \cdot \frac{\partial S_{GBP}}{\partial \rho_{EUR,USD}} + \frac{\partial C}{\partial FX_{USD/EUR}} \cdot \frac{\partial FX_{USD/EUR}}{\partial \rho_{EUR,USD}}
\end{aligned}
\end{equation}

\noindent Simplifying further, we get:
\begin{equation}
\begin{aligned}
\text{Cora}_{\text{GBP,USD}} &= \frac{\partial C}{\partial S_{GBP}} \cdot \frac{\partial S_{GBP}}{\partial \rho_{GBP,USD}} + \frac{\partial C}{\partial S_{USD}} \cdot \frac{\partial S_{USD}}{\partial \rho_{GBP,USD}} \\
&+ \frac{\partial C}{\partial S_{EUR}} \cdot \frac{\partial S_{EUR}}{\partial \rho_{GBP,USD}} 
\end{aligned}
\end{equation}

\begin{equation}
\begin{aligned}
\text{Cora}_{\text{EUR,USD}} &= \frac{\partial C}{\partial S_{EUR}} \cdot \frac{\partial S_{EUR}}{\partial \rho_{EUR,USD}} + \frac{\partial C}{\partial S_{USD}} \cdot \frac{\partial S_{USD}}{\partial \rho_{EUR,USD}} \\
&+ \frac{\partial C}{\partial S_{GBP}} \cdot \frac{\partial S_{GBP}}{\partial \rho_{EUR,USD}} 
\end{aligned}
\end{equation}

\noindent For the second partial derivatives:
\begin{align}
\frac{\partial^2 C}{\partial \rho_{GBP,USD}^2} &= \frac{\partial}{\partial \rho_{GBP,USD}} \left( \frac{\partial C}{\partial S_{GBP}} \cdot \frac{\partial S_{GBP}}{\partial \rho_{GBP,USD}} + \frac{\partial C}{\partial S_{USD}} \cdot \frac{\partial S_{USD}}{\partial \rho_{GBP,USD}} \right. \nonumber\\ 
&\quad \left. + \frac{\partial C}{\partial S_{EUR}} \cdot \frac{\partial S_{EUR}}{\partial \rho_{GBP,USD}} \right) \nonumber\\
&= \frac{\partial}{\partial \rho_{GBP,USD}} \left( \frac{\partial C}{\partial S_{GBP}} \right) \cdot \frac{\partial S_{GBP}}{\partial \rho_{GBP,USD}} + \frac{\partial C}{\partial S_{GBP}} \cdot \frac{\partial}{\partial \rho_{GBP,USD}} \left( \frac{\partial S_{GBP}}{\partial \rho_{GBP,USD}} \right) \nonumber\\ 
&+ \frac{\partial}{\partial \rho_{GBP,USD}} \left( \frac{\partial C}{\partial S_{USD}} \right) \cdot \frac{\partial S_{USD}}{\partial \rho_{GBP,USD}} + \frac{\partial C}{\partial S_{USD}} \cdot \frac{\partial}{\partial \rho_{GBP,USD}} \left( \frac{\partial S_{USD}}{\partial \rho_{GBP,USD}} \right) \nonumber\\ 
&+ \frac{\partial}{\partial \rho_{GBP,USD}} \left( \frac{\partial C}{\partial S_{EUR}} \right) \cdot \frac{\partial S_{EUR}}{\partial \rho_{GBP,USD}} + \frac{\partial C}{\partial S_{EUR}} \cdot \frac{\partial}{\partial \rho_{GBP,USD}} \left( \frac{\partial S_{EUR}}{\partial \rho_{GBP,USD}} \right) 
\end{align}

\noindent Similarly, for \(\rho_{EUR,USD}\):
\begin{align}
\frac{\partial^2 C}{\partial \rho_{EUR,USD}^2} &= \frac{\partial}{\partial \rho_{EUR,USD}} \left( \frac{\partial C}{\partial S_{EUR}} \cdot \frac{\partial S_{EUR}}{\partial \rho_{EUR,USD}} + \frac{\partial C}{\partial S_{USD}} \cdot \frac{\partial S_{USD}}{\partial \rho_{EUR,USD}} \right. \nonumber\\
&\quad \left. + \frac{\partial C}{\partial S_{GBP}} \cdot \frac{\partial S_{GBP}}{\partial \rho_{EUR,USD}} \right) \nonumber\\
&= \frac{\partial}{\partial \rho_{EUR,USD}} \left( \frac{\partial C}{\partial S_{EUR}} \right) \cdot \frac{\partial S_{EUR}}{\partial \rho_{EUR,USD}} + \frac{\partial C}{\partial S_{EUR}} \cdot \frac{\partial}{\partial \rho_{EUR,USD}} \left( \frac{\partial S_{EUR}}{\partial \rho_{EUR,USD}} \right) \nonumber\\ 
&+ \frac{\partial}{\partial \rho_{EUR,USD}} \left( \frac{\partial C}{\partial S_{USD}} \right) \cdot \frac{\partial S_{USD}}{\partial \rho_{EUR,USD}} + \frac{\partial C}{\partial S_{USD}} \cdot \frac{\partial}{\partial \rho_{EUR,USD}} \left( \frac{\partial S_{USD}}{\partial \rho_{EUR,USD}} \right) \nonumber\\ 
&+ \frac{\partial}{\partial \rho_{EUR,USD}} \left( \frac{\partial C}{\partial S_{GBP}} \right) \cdot \frac{\partial S_{GBP}}{\partial \rho_{EUR,USD}} + \frac{\partial C}{\partial S_{GBP}} \cdot \frac{\partial}{\partial \rho_{EUR,USD}} \left( \frac{\partial S_{GBP}}{\partial \rho_{EUR,USD}} \right) 
\end{align}

\noindent Therefore, we get:
\begin{align}
\text{Gora}_{\text{GBP,USD}} &= \frac{\partial^2 C}{\partial S_{GBP} \partial \rho_{GBP,USD}} \cdot \frac{\partial S_{GBP}}{\partial \rho_{GBP,USD}} + \frac{\partial C}{\partial S_{GBP}} \cdot \frac{\partial^2 S_{GBP}}{\partial \rho_{GBP,USD}^2} \nonumber\\
&+ \frac{\partial^2 C}{\partial S_{USD} \partial \rho_{GBP,USD}} \cdot \frac{\partial S_{USD}}{\partial \rho_{GBP,USD}} + \frac{\partial C}{\partial S_{USD}} \cdot \frac{\partial^2 S_{USD}}{\partial \rho_{GBP,USD}^2} \nonumber\\
&+ \frac{\partial^2 C}{\partial S_{EUR} \partial \rho_{GBP,USD}} \cdot \frac{\partial S_{EUR}}{\partial \rho_{GBP,USD}} + \frac{\partial C}{\partial S_{EUR}} \cdot \frac{\partial^2 S_{EUR}}{\partial \rho_{GBP,USD}^2} \nonumber\\
\end{align}

\begin{align}
\text{Gora}_{\text{EUR,USD}} &= \frac{\partial^2 C}{\partial S_{EUR} \partial \rho_{EUR,USD}} \cdot \frac{\partial S_{EUR}}{\partial \rho_{EUR,USD}} + \frac{\partial C}{\partial S_{EUR}} \cdot \frac{\partial^2 S_{EUR}}{\partial \rho_{EUR,USD}^2} \nonumber\\
&+ \frac{\partial^2 C}{\partial S_{USD} \partial \rho_{EUR,USD}} \cdot \frac{\partial S_{USD}}{\partial \rho_{EUR,USD}} + \frac{\partial C}{\partial S_{USD}} \cdot \frac{\partial^2 S_{USD}}{\partial \rho_{EUR,USD}^2} \nonumber\\
&+ \frac{\partial^2 C}{\partial S_{GBP} \partial \rho_{EUR,USD}} \cdot \frac{\partial S_{GBP}}{\partial \rho_{EUR,USD}} + \frac{\partial C}{\partial S_{GBP}} \cdot \frac{\partial^2 S_{GBP}}{\partial \rho_{EUR,USD}^2} \nonumber\\
\end{align}
\\

\section{CONCLUSION}
Ultimately, we conclude that to accurately price and hedge multi-asset foreign equity quanto options, it is essential to incorporate stochastic correlations alongside stochastic volatilities and the appropriate modeling of stochastic FX rates. The payoffs of our Quanto call options incorporate multiple correlated underlying assets weighted by the FX rate values (which are uncorrelated in our paper). This makes accurate modeling of the FX rates and stochastic correlation between the BMs of the underlying assets essential for achieving optimal performance in MC simulations. By systematically testing all combinations of choices for a varied selection of SDEs for SC, SV, and SER, we identify the most effective configuration of our model. Overall, the combination of GARCH-Jump SV, OU FX rates, Weibull SC, and the Milstein or Runge-Kutta discretization scheme consistently performs well across both cases of the option and start dates we tested. We also find that incorporating mean reversion into stochastic correlation or stochastic FX rate modeling is beneficial for MC simulation pricing. Specifically, it seems that incorporating mean reversion into stochastic correlation models is beneficial not only to ensure simulated correlation paths stay within the realistic range [-1, 1] but also since the motions of correlations between assets observed in the market demonstrate this property. 
\\

\noindent Moreover, hedging correlation risks is a crucial aspect of using multi-asset Quanto options effectively in practice. We derive formulas for Cora and Gora of our Quanto options in terms of partial derivatives. Our derived Cora and Gora expressions can be made even more explicit by evaluating the partial derivatives for specific choices of SV and SC of the underlying assets, which allows for efficient hedging of correlation risk for both cases of the options. Finally, based on our findings, we conjecture that the choice to model volatility as stochastic (vs. constant) is relatively more significant for pricing accuracy than modeling correlations as stochastic. Something we want to explore further is whether there is some number (of correlated processes acting as the underlying assets for the option) for which this relative importance of modeling volatility or correlation as stochastic is reversed. Increasing the number of correlated assets should increase the correlation risk associated with the option, and hence, this should increase the importance of how we model correlations. However, it is not clear whether there are some conditions for the option's setup for which modeling correlations as stochastic have more of an effect on pricing accuracy than modeling volatilities as stochastic. Hence, it would be beneficial to study further how the relative effectiveness of these choices changes with different market conditions.

\section*{Future Work}
The next step in our research is to extend our framework to handle payoffs involving three or more assets without necessarily including an asset in the investor's domestic currency. The primary challenge of this extension is the increased number of stochastic differential equations (SDEs) that must be discretized and simulated, which grows quickly with the addition of more currencies and foreign equity indices. Another potential consideration for future work is introducing jumps into the stochastic correlation processes as we did with volatility (Bates and GARCH-Jump models). Whilst we do not test such models in this paper, sudden moves in correlations are also feasible in some market conditions, so this should be studied. In addition, modeling volatilities as driven by fractional Itô processes ('fractional stochastic' or 'rough' volatility models) is worth exploring (\cite{AlForaih2023}, \cite{Chong2024}, \cite{Jacquier2023}, \cite{Horst2023}, \cite{Fukasawa2024}, \cite{AbiJaber2019}, \cite{Fukasawa2023}, \cite{Kim2024}, \cite{Garnier2017}, \cite{Bourgey2022}, \cite{Cont2023}, \cite{Comte2012}, \cite{Alos2017}, \cite{Alos2018}, \cite{Shi2024}). As proposed by Comte and Renault (1998) in their development of the Fractional Stochastic Volatility (FSV) model \cite{Comte1998}, fractional Brownian motion (fBM) with Hurst parameter \( H > \frac{1}{2} \) can be used to better capture the long-memory property of volatility. The fractional stochastic differential equation considered in their model is given by:
\begin{equation}
dX_t = -\kappa X_t dt + \sigma dB_t^\alpha \quad x_0 = 0, \kappa > 0, \alpha := H - \frac{1}{2}, 0 < \alpha < \frac{1}{2}  
\end{equation}

The FSV model utilizes fractional Brownian motion \( B_t^H \) to incorporate long-memory effects, where the increments of \( B_t^H \) are positively correlated when \( H > \frac{1}{2} \). This allows the model to capture the mean-reverting nature of volatility without explicit mean-reversion terms, as explained by Shi et al. in their paper \textit{Fractional Stochastic Volatility Model (2021)} \cite{Shi2021}. The significance of the Hurst exponent \( H \) lies in its ability to describe the roughness or smoothness of the volatility paths, with \( H < \frac{1}{2} \) indicating roughness and \( H > \frac{1}{2} \) indicating smoothness. This approach has the advantage of supposedly aligning better with observed market volatilities compared to traditional models \cite{Shi2021}. Additionally, there is potential to extend this framework to model stochastic correlation, allowing for both fractional stochastic volatility and fractional stochastic correlation. As shown in Figure 32, for shorter rolling windows the observed correlations look like they could potentially be modeled more effectively by a fractional stochastic process.
\\

\section*{List of Figures}
\begin{figure}[H]
    \centering
    \includegraphics[width=0.65\linewidth]{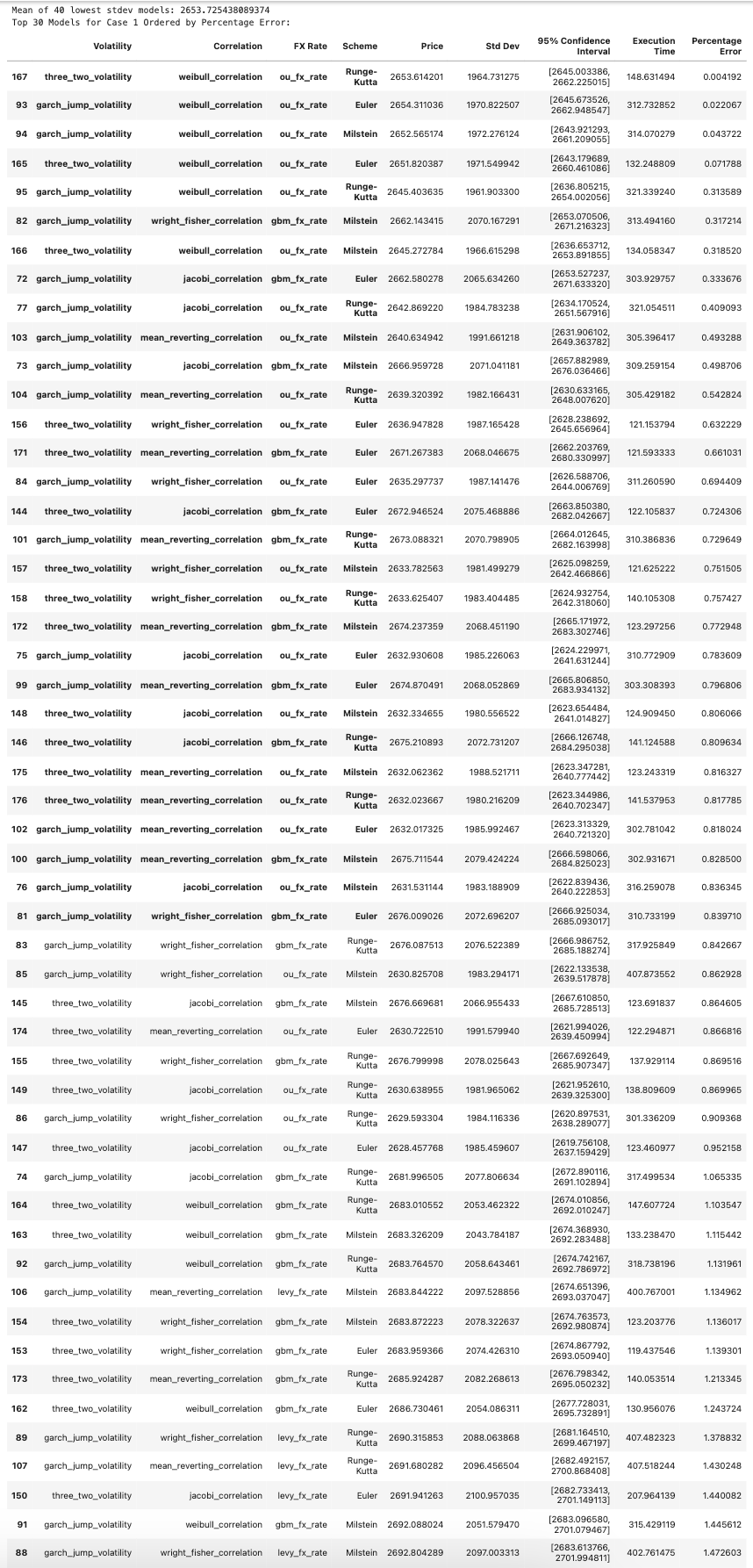} 
    \caption{Performance of MC Simulation by Percentage Error, 2021 Start, Case 1, Part 1}
    \label{fig:Case_1_1st_2021}
\end{figure}

\begin{figure}[H]
    \centering
    \includegraphics[width=0.70\linewidth]{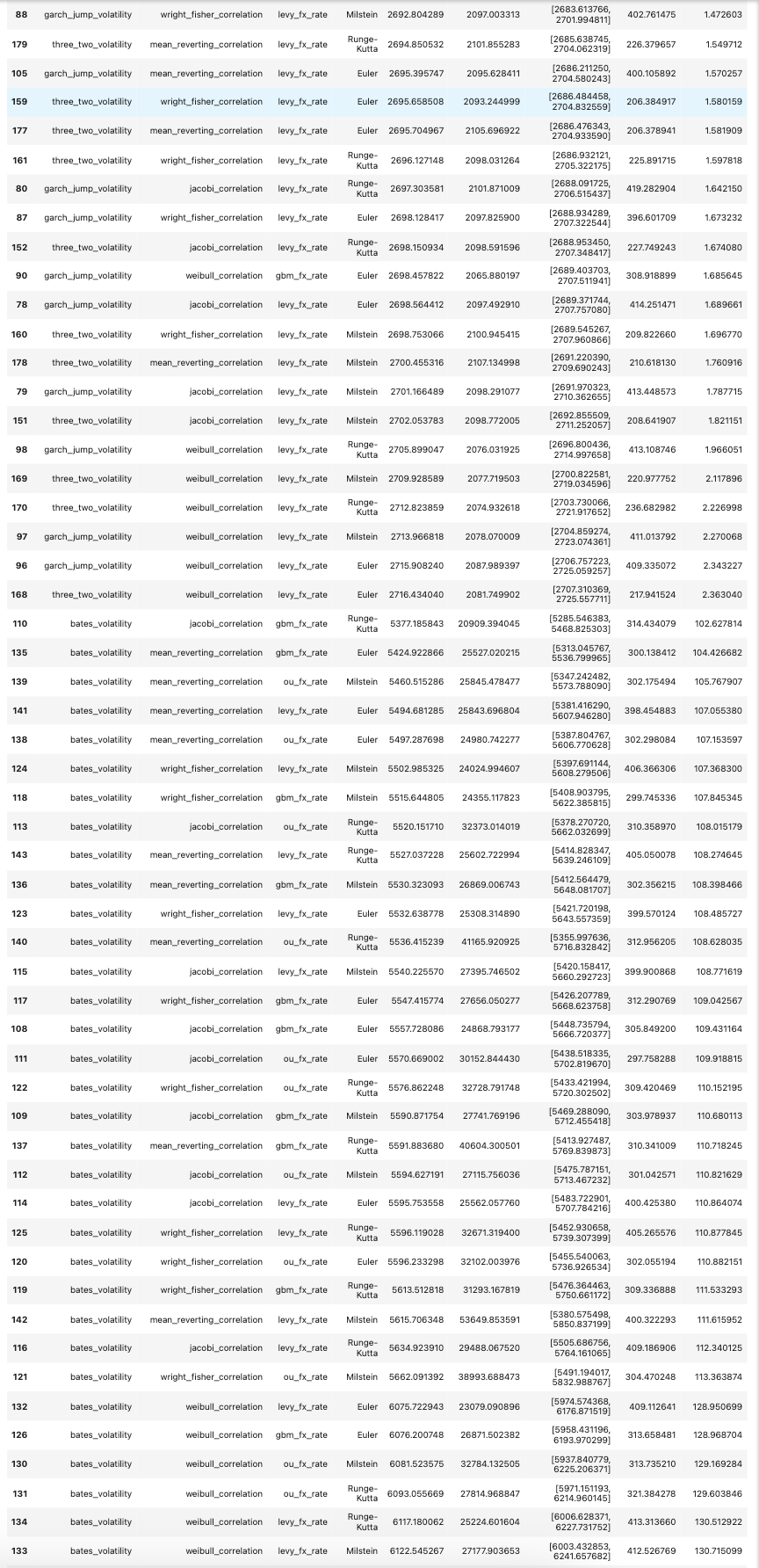} 
    \caption{Performance of MC Simulation by Percentage Error, 2021 Start, Case 1, Part 2}
    \label{fig:Case_1_2nd_2021}
\end{figure}

\begin{figure}[H]
    \centering
    \includegraphics[width=0.85\linewidth]{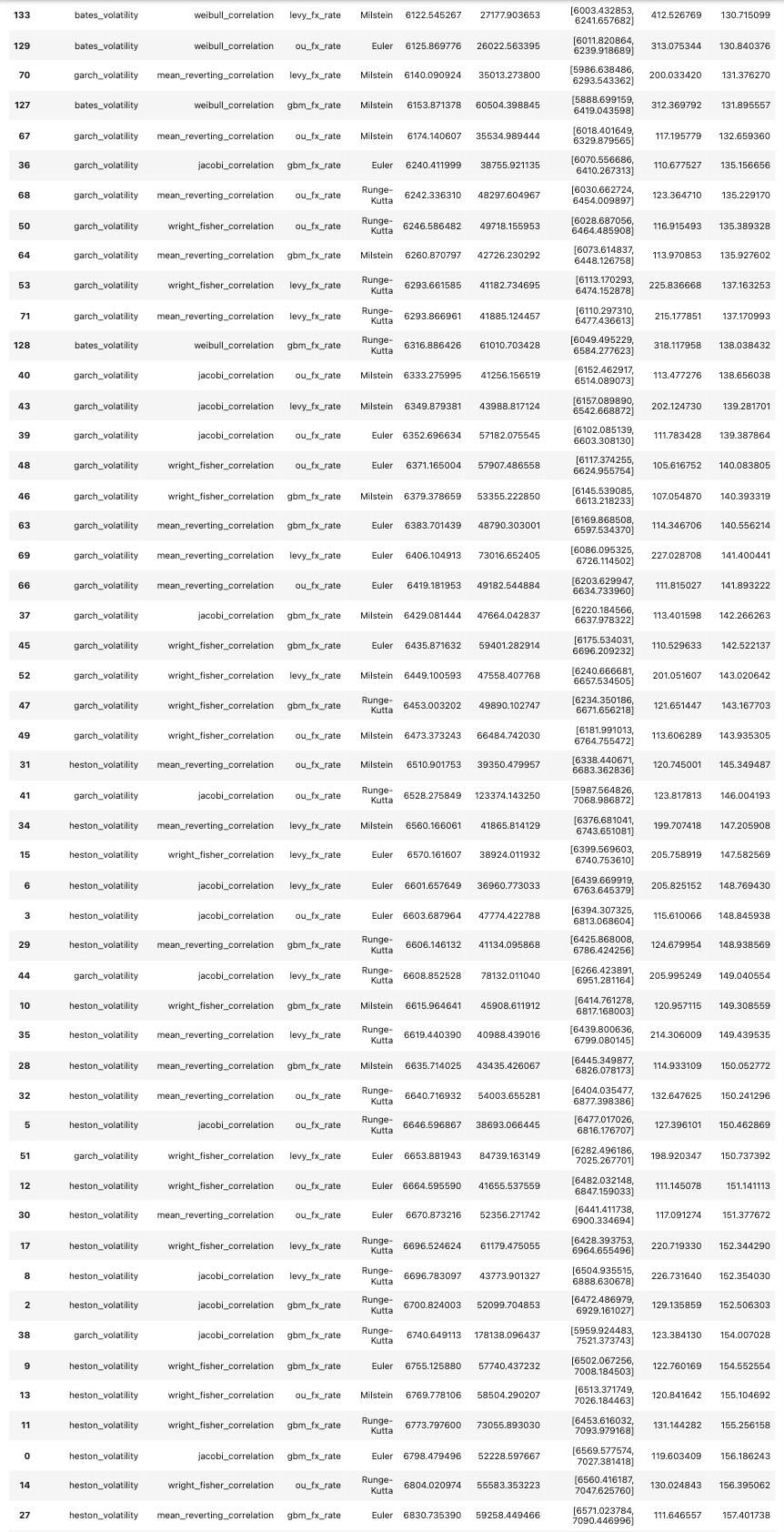} 
    \caption{Performance of MC Simulation by Percentage Error, 2021 Start, Case 1, Part 3}
    \label{fig:Case_1_3rd_2021}
\end{figure}

\begin{figure}[H]
    \centering
    \includegraphics[width=0.85\linewidth]{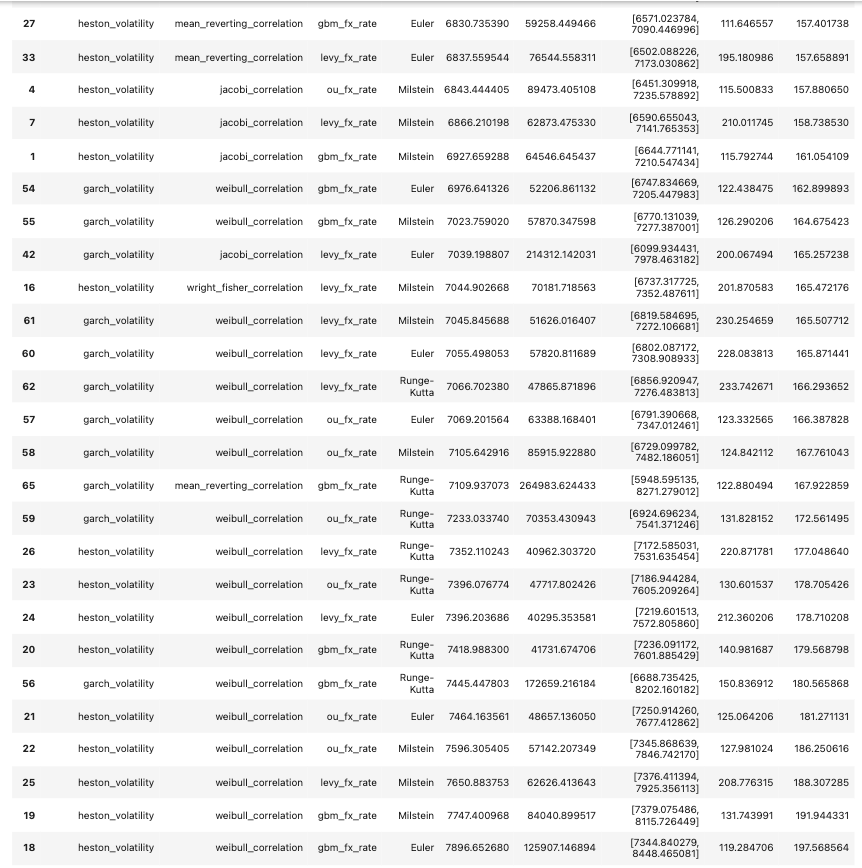} 
    \caption{Performance of MC Simulation by Percentage Error, 2021 Start, Case 1, Part 4}
    \label{fig:Case_1_4th_2021}
\end{figure}

\begin{figure}[H]
    \centering
    \includegraphics[width=0.85\linewidth]{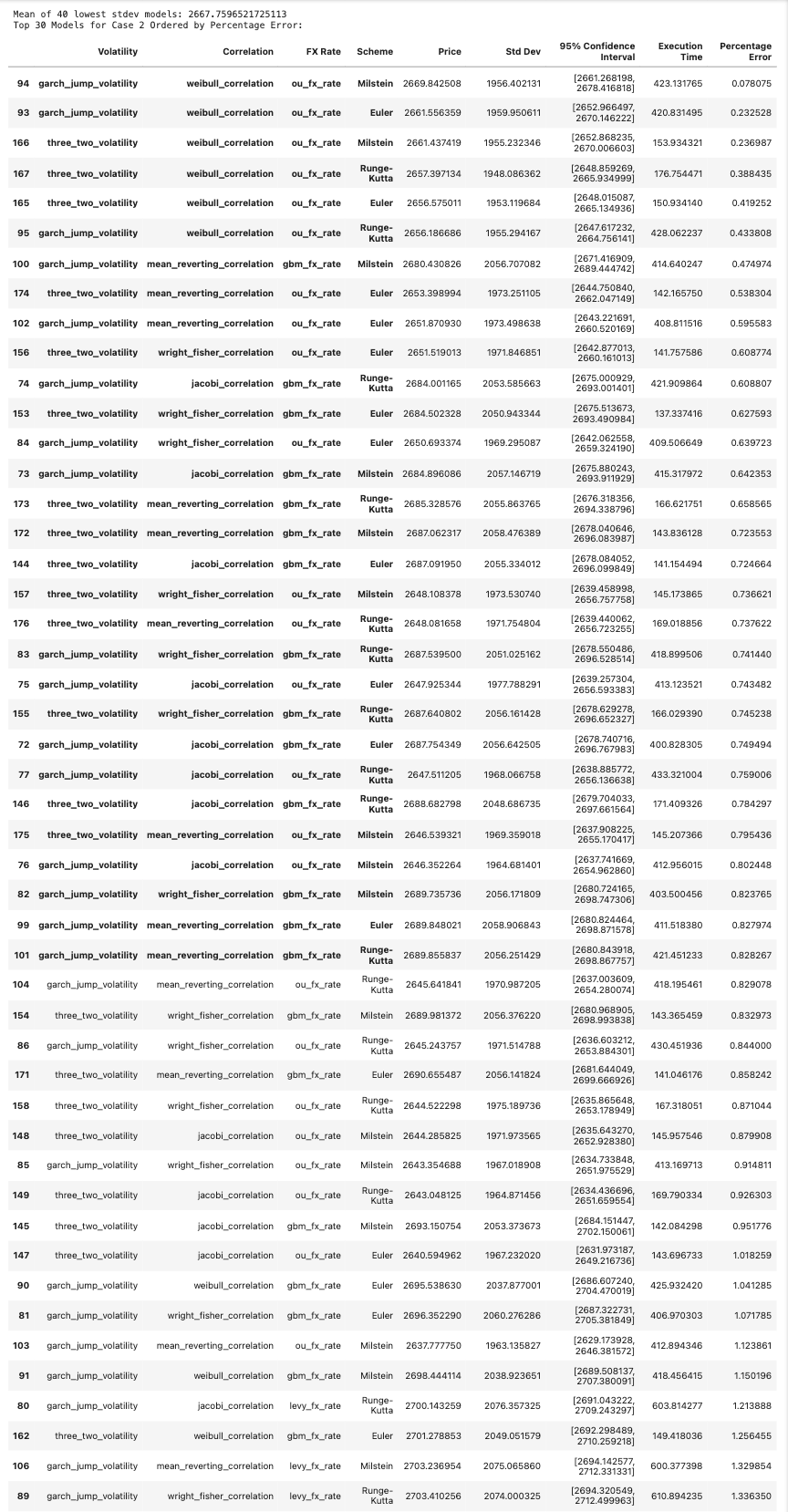} 
    \caption{Performance of MC Simulation by Percentage Error, 2021 Start, Case 2, Part 1}
    \label{fig:Case_2_1st_2021}
\end{figure}

\begin{figure}[H]
    \centering
    \includegraphics[width=0.85\linewidth]{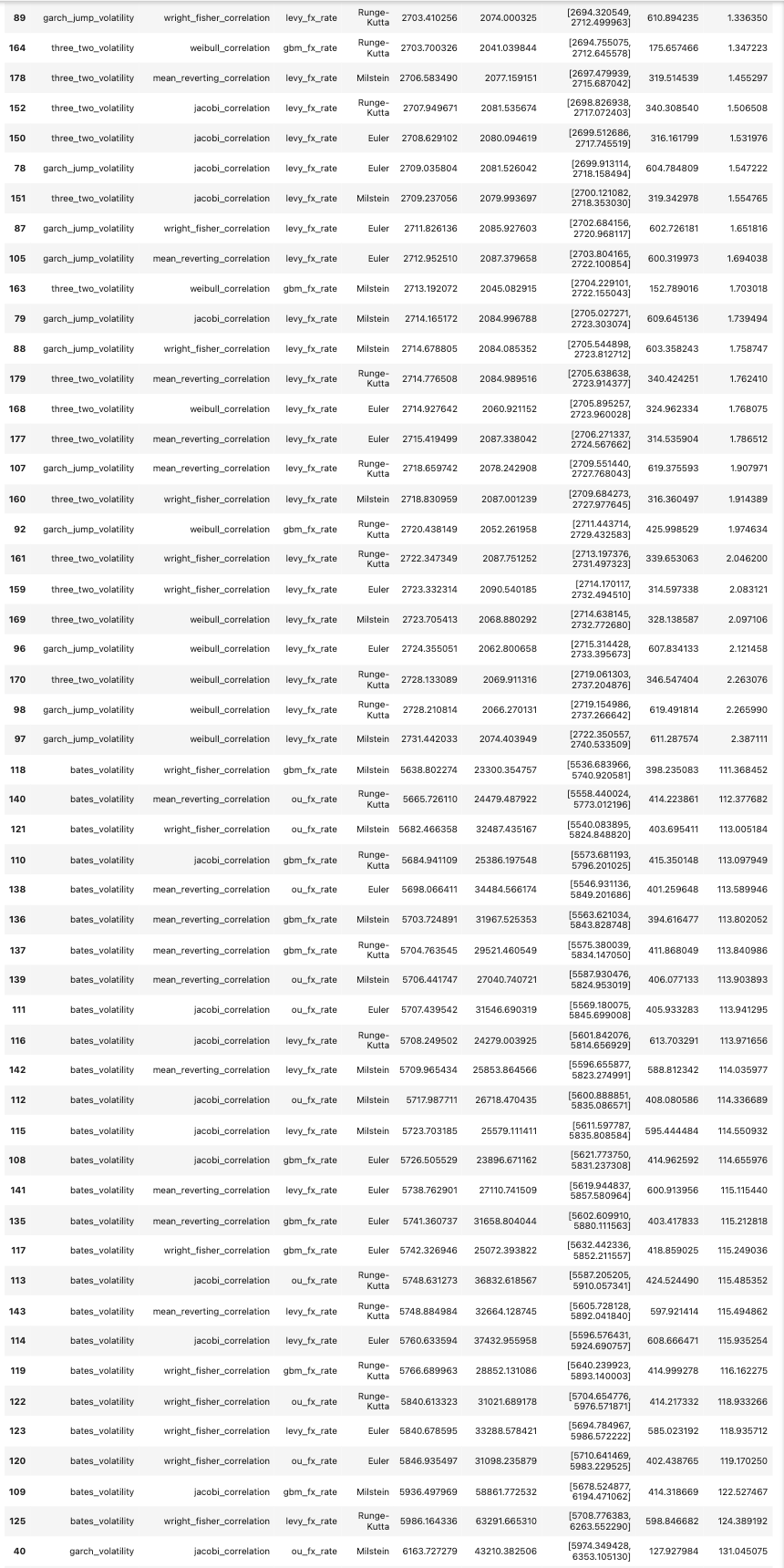} 
    \caption{Performance of MC Simulation by Percentage Error, 2021 Start, Case 2, Part 2}
    \label{fig:Case_2_2nd_2021}
\end{figure}

\begin{figure}[H]
    \centering
    \includegraphics[width=0.85\linewidth]{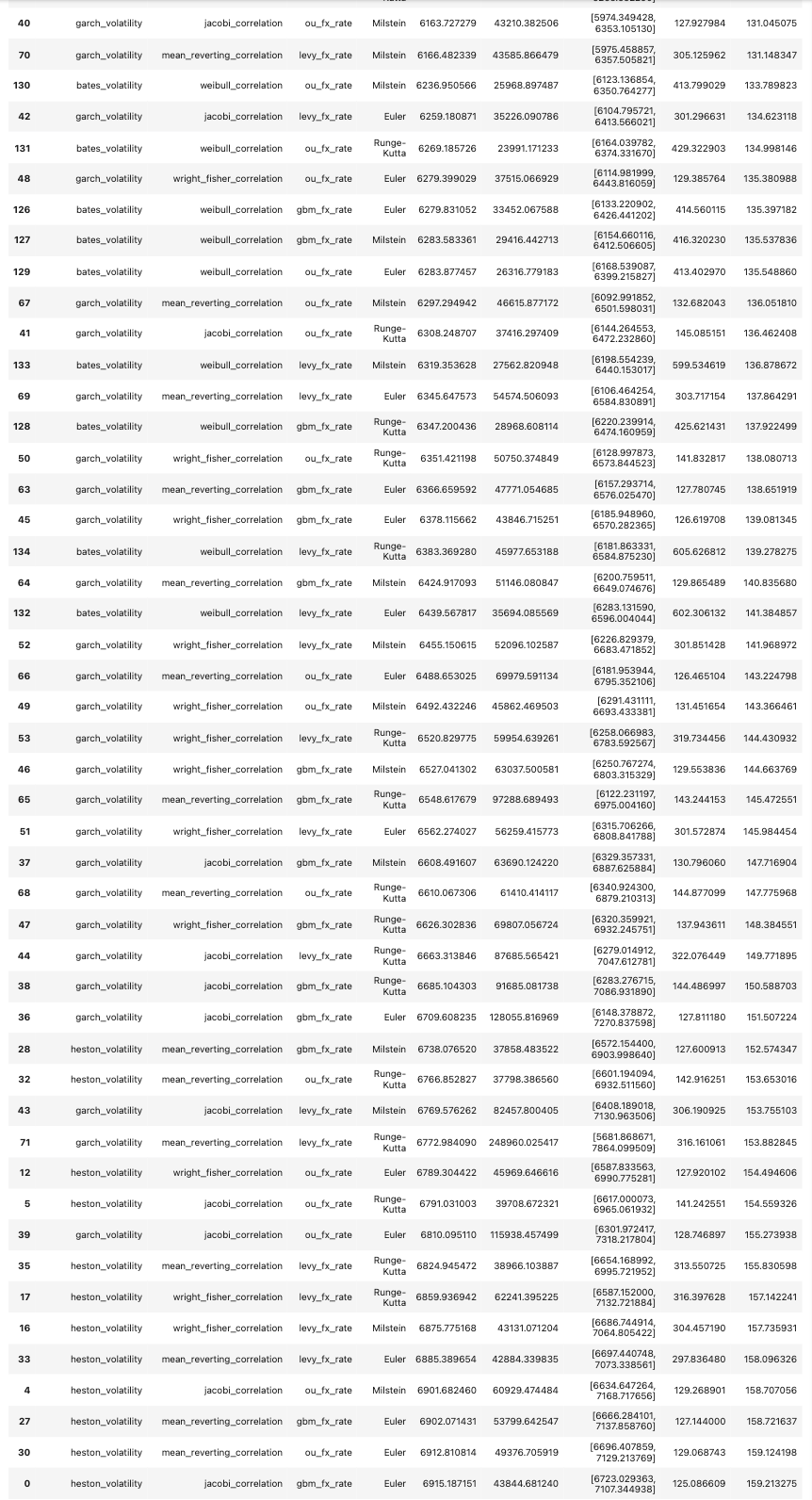} 
    \caption{Performance of MC Simulation by Percentage Error, 2021 Start, Case 2, Part 3}
    \label{fig:Case_2_3rd_2021}
\end{figure}

\begin{figure}[H]
    \centering
    \includegraphics[width=0.85\linewidth]{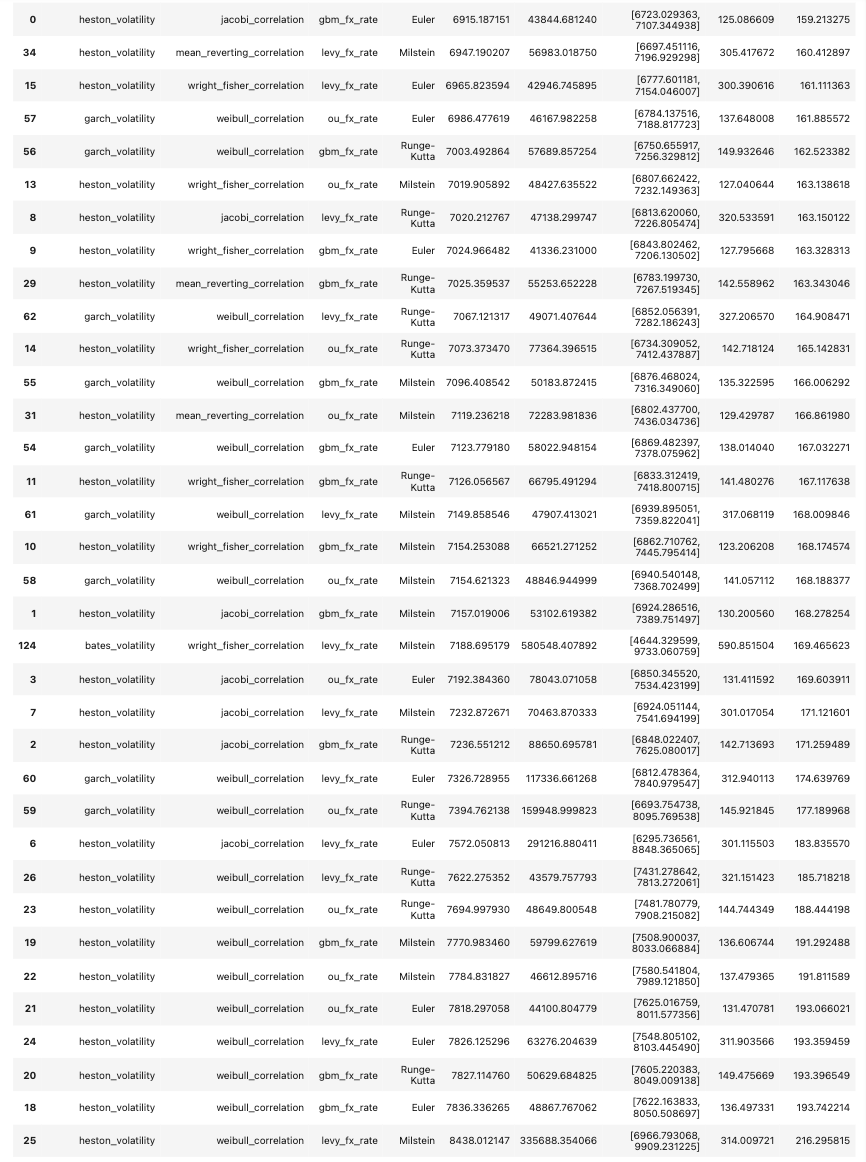} 
    \caption{Performance of MC Simulation by Percentage Error, 2021 Start, Case 2, Part 4}
    \label{fig:Case_2_4th_2021}
\end{figure}

\begin{figure}[H]
    \centering
    \includegraphics[width=0.85\linewidth]{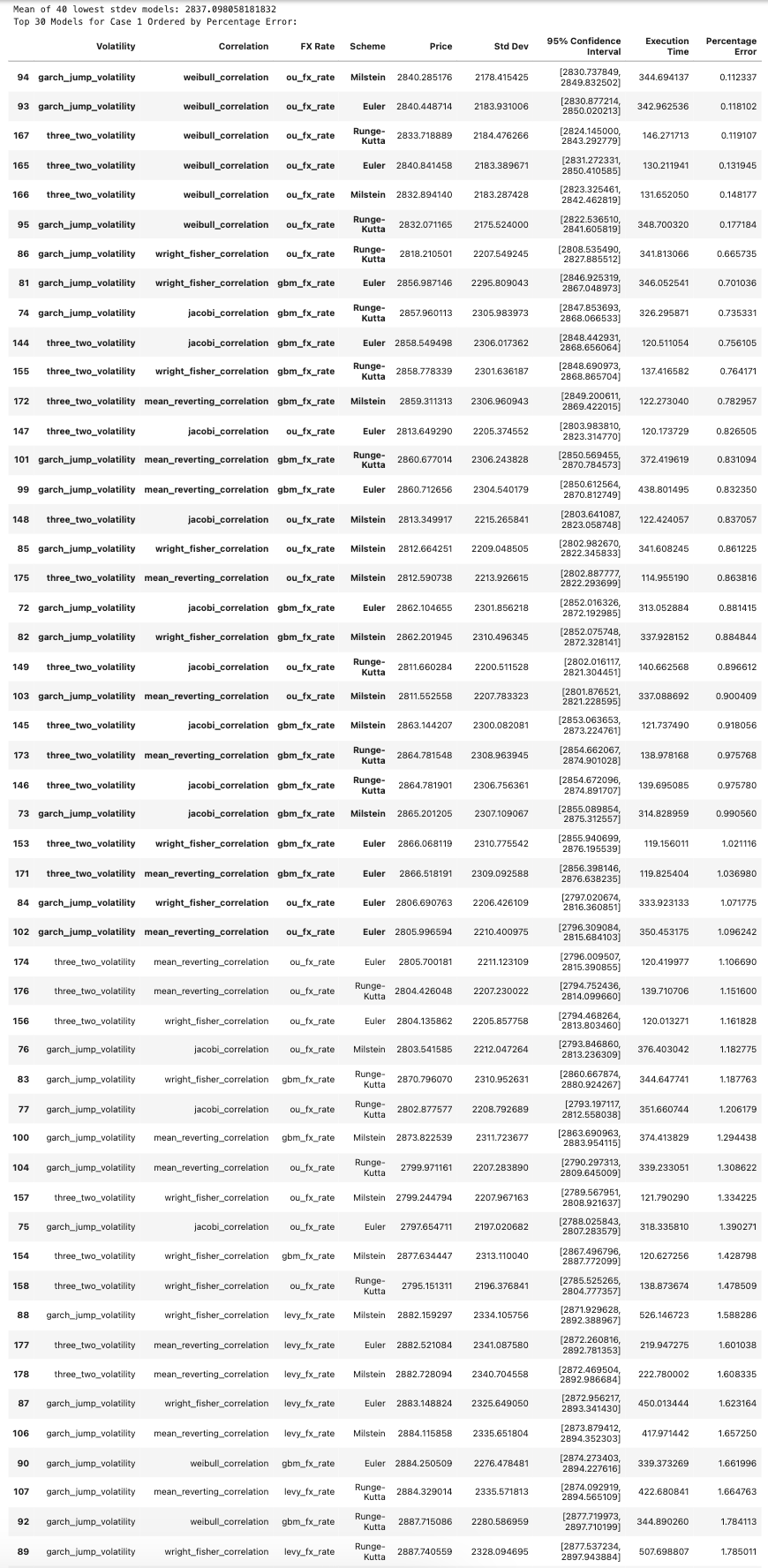} 
    \caption{Performance of MC Simulation by Percentage Error, 2022 Start, Case 1, Part 1}
    \label{fig:Case_1_1st_2022}
\end{figure}

\begin{figure}[H]
    \centering
    \includegraphics[width=0.80\linewidth]{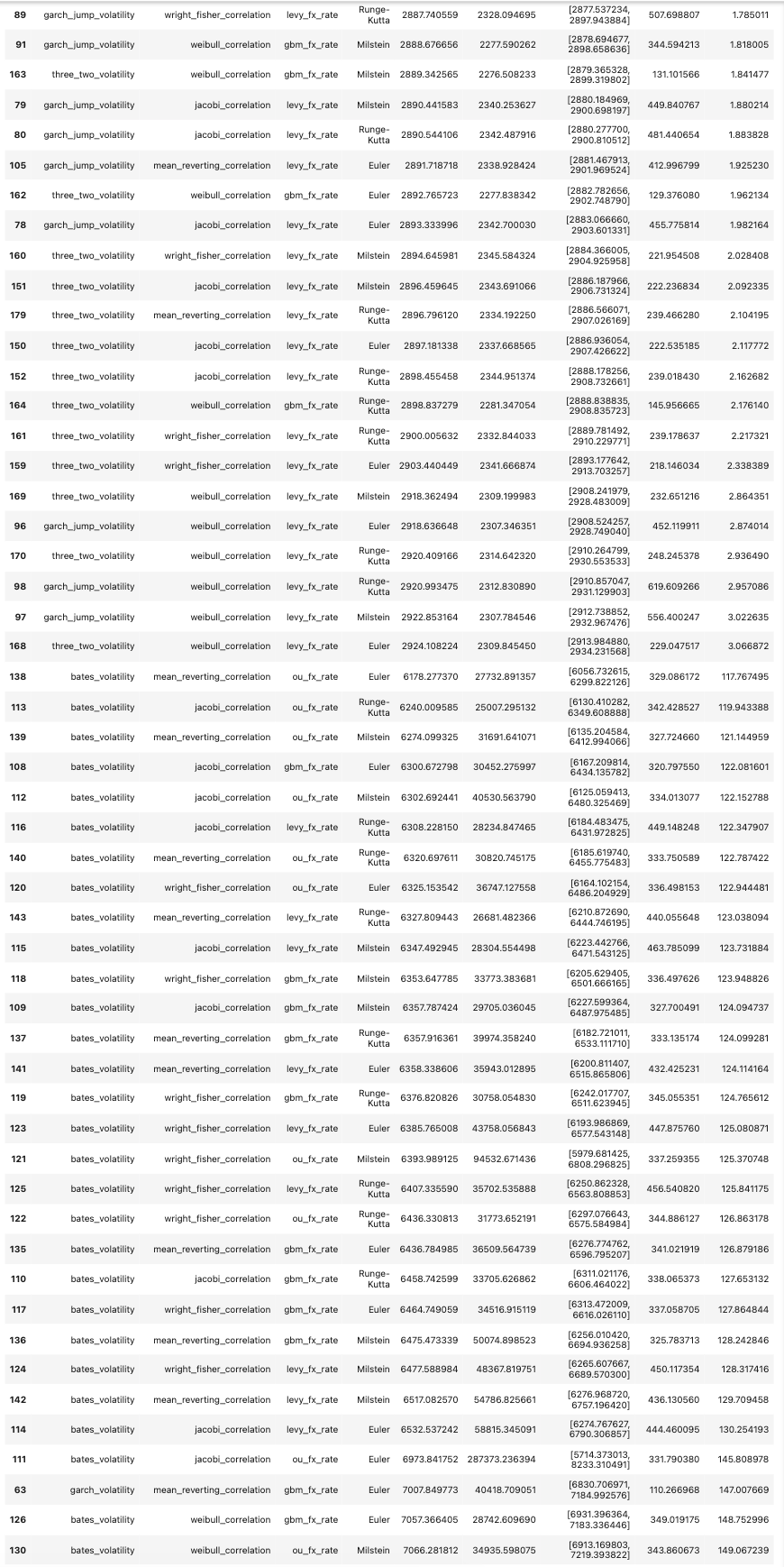} 
    \caption{Performance of MC Simulation by Percentage Error, 2022 Start, Case 1, Part 2}
    \label{fig:Case_1_2nd_2022}
\end{figure}

\begin{figure}[H]
    \centering
    \includegraphics[width=0.80\linewidth]{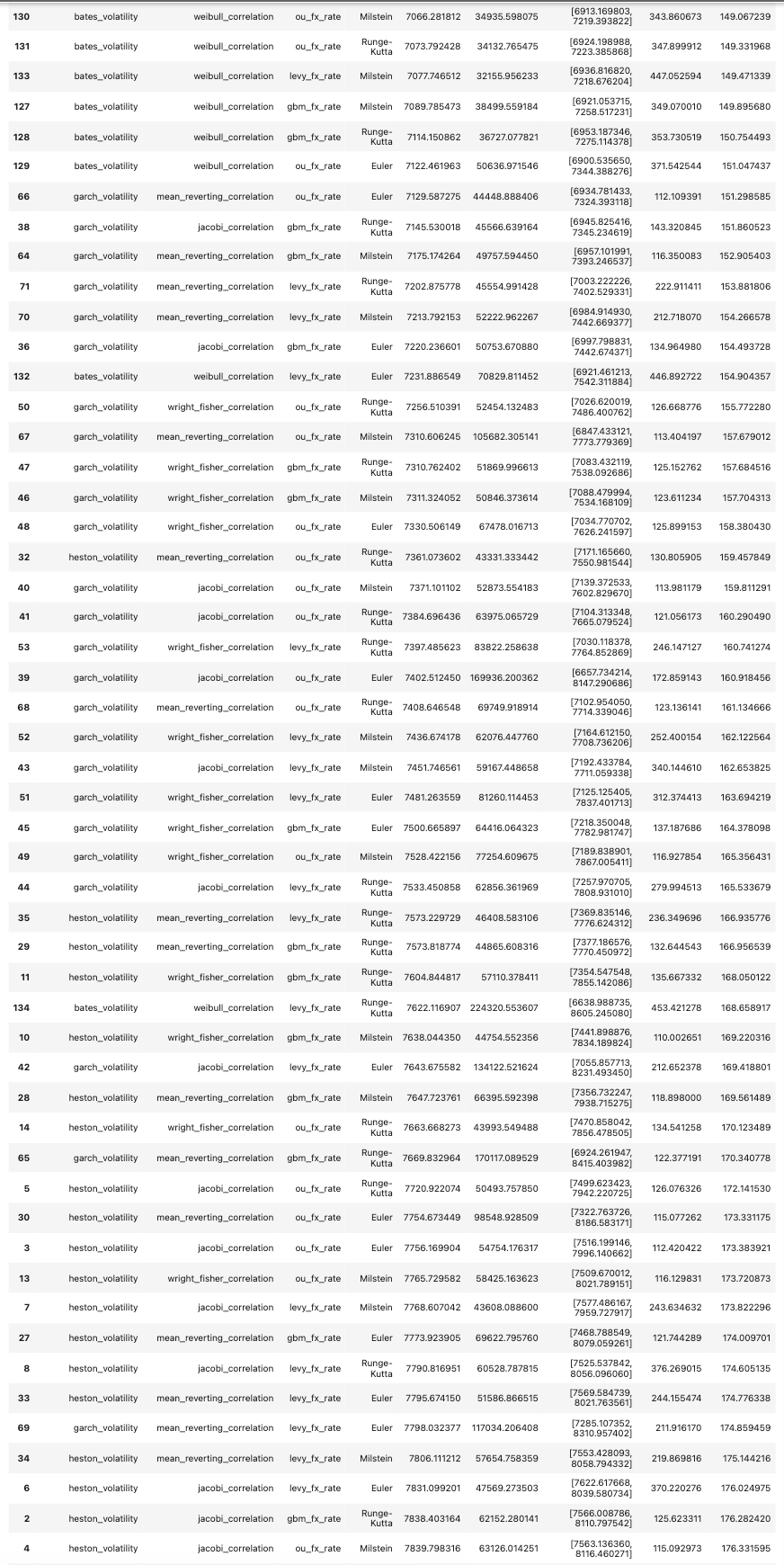} 
    \caption{Performance of MC Simulation by Percentage Error, 2022 Start, Case 1, Part 3}
    \label{fig:Case_1_3rd_2022}
\end{figure}

\begin{figure}[H]
    \centering
    \includegraphics[width=0.85\linewidth]{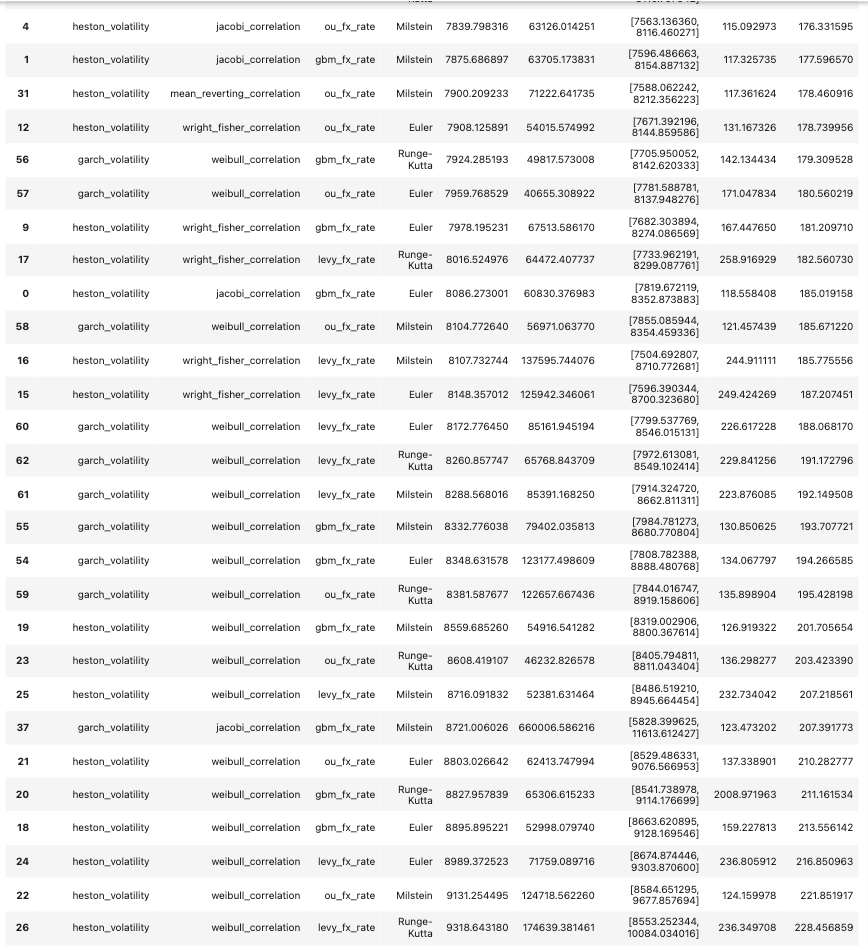} 
    \caption{Performance of MC Simulation by Percentage Error, 2022 Start, Case 1, Part 4}
    \label{fig:Case_1_4th_2022}
\end{figure}

\begin{figure}[H]
    \centering
    \includegraphics[width=0.85\linewidth]{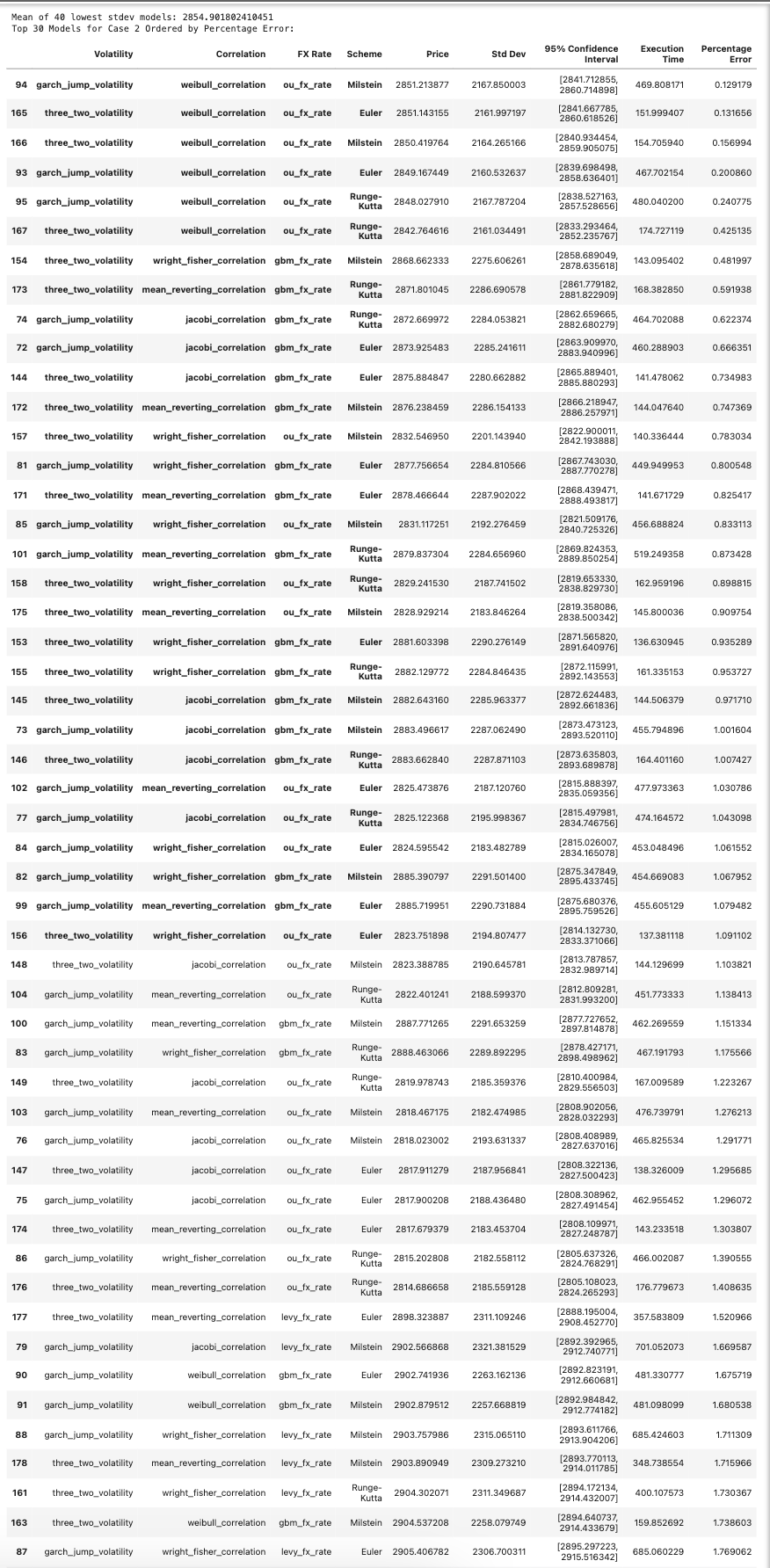} 
    \caption{Performance of MC Simulation by Percentage Error, 2022 Start, Case 2, Part 1}
    \label{fig:Case_2_1st_2022}
\end{figure}

\begin{figure}[H]
    \centering
    \includegraphics[width=0.80\linewidth]{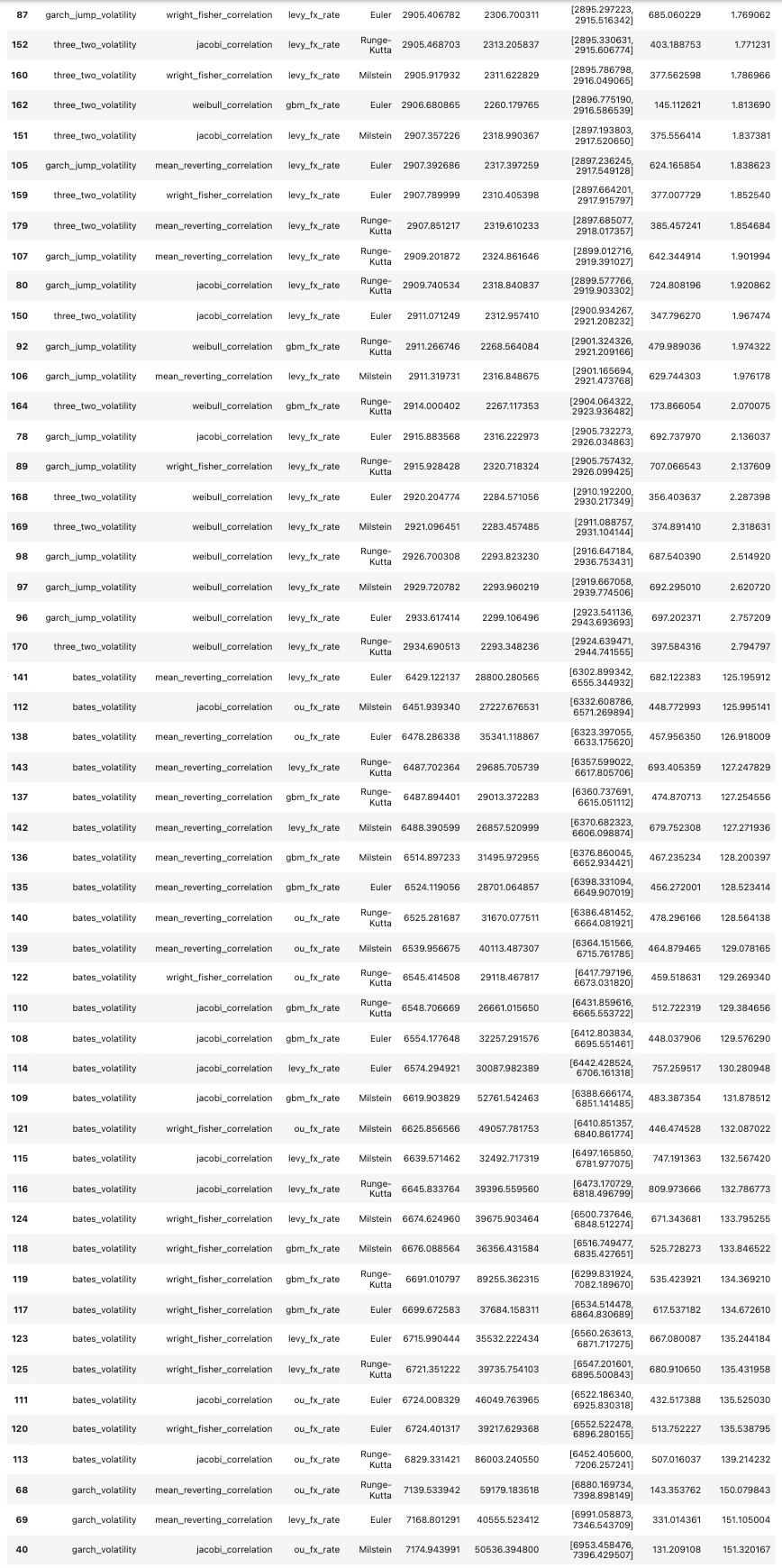} 
    \caption{Performance of MC Simulation by Percentage Error, 2022 Start, Case 2, Part 2}
    \label{fig:Case_2_2nd_2022}
\end{figure}

\begin{figure}[H]
    \centering
    \includegraphics[width=0.80\linewidth]{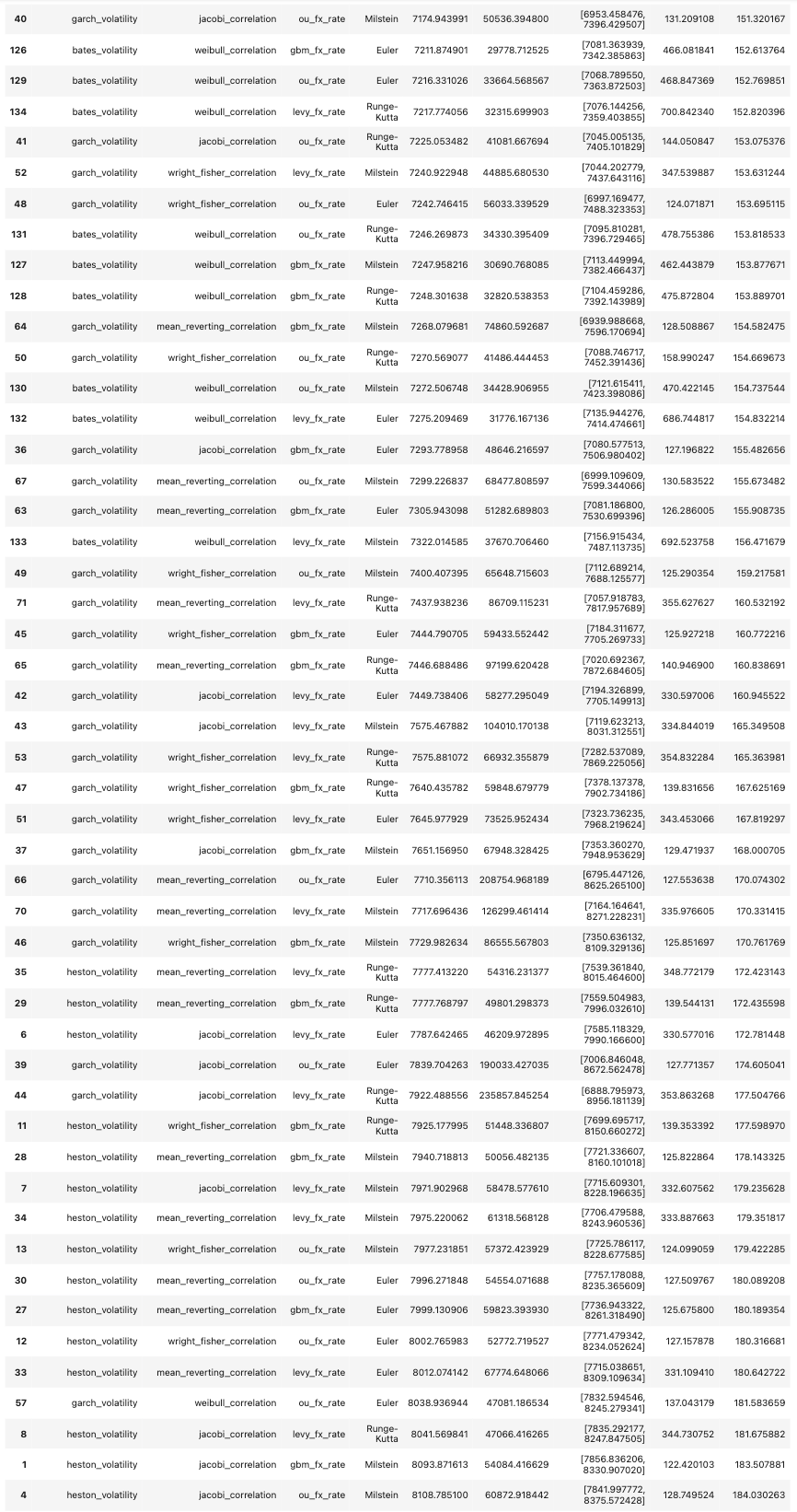} 
    \caption{Performance of MC Simulation by Percentage Error, 2022 Start, Case 2, Part 3}
    \label{fig:Case_2_3rd_2022}
\end{figure}

\begin{figure}[H]
    \centering
    \includegraphics[width=0.85\linewidth]{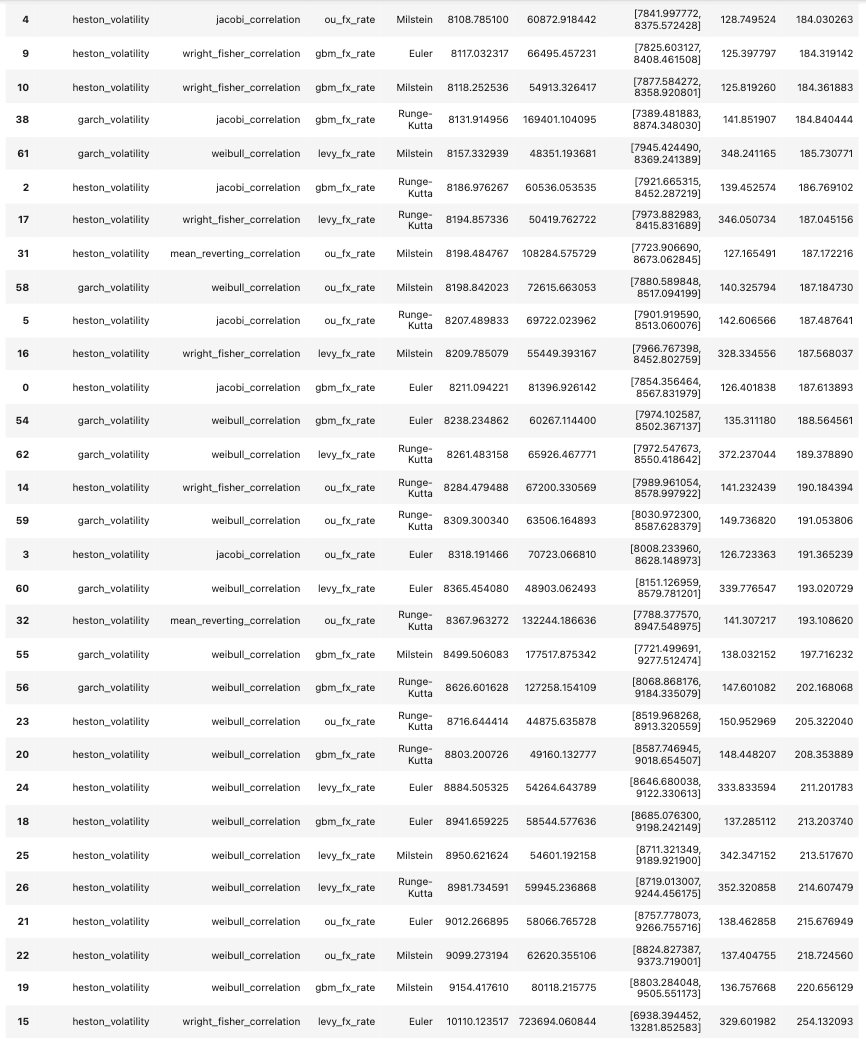} 
    \caption{Performance of MC Simulation by Percentage Error, 2022 Start, Case 2, Part 4}
    \label{fig:Case_2_4th_2022}
\end{figure}

\begin{figure}[H]
    \centering
    \includegraphics[width=0.85\linewidth]{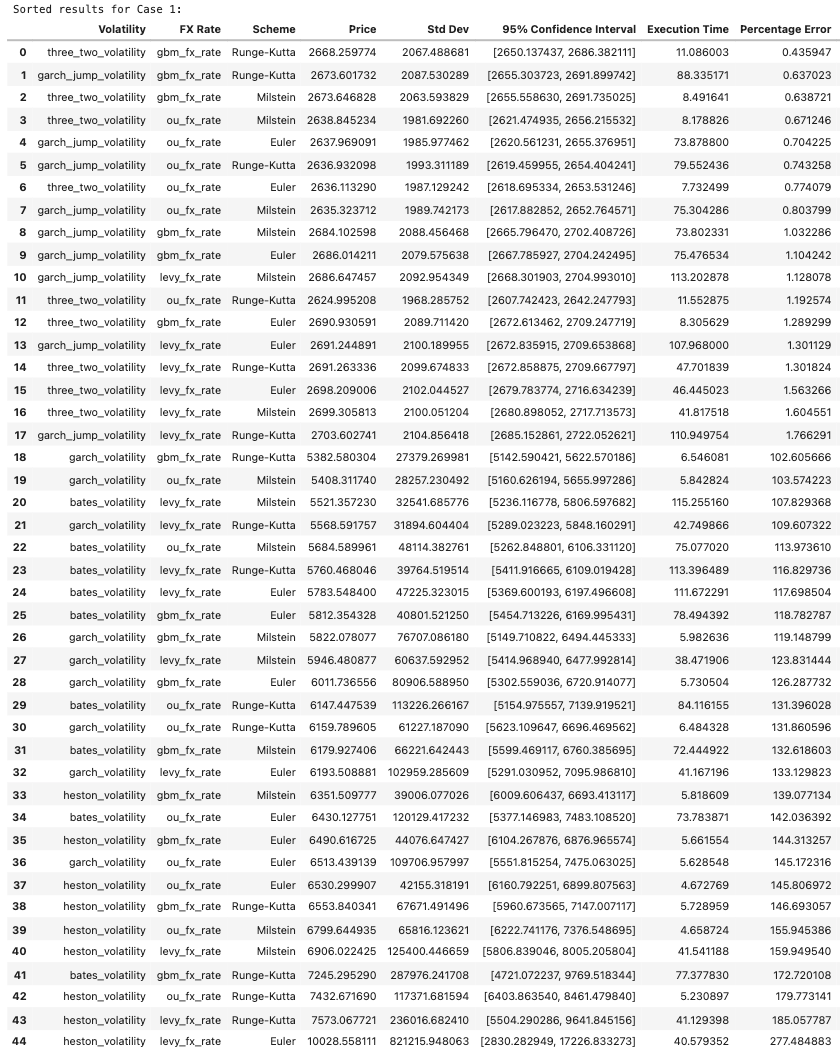} 
    \caption{Performance of MC Simulation with Constant Correlation by Percentage Error, 2021 Start, Case 1}
    \label{fig:MC_Sim_constant_corr_by_percentage_error_2021_start_case_1}
\end{figure}

\begin{figure}[H]
    \centering
    \includegraphics[width=0.85\linewidth]{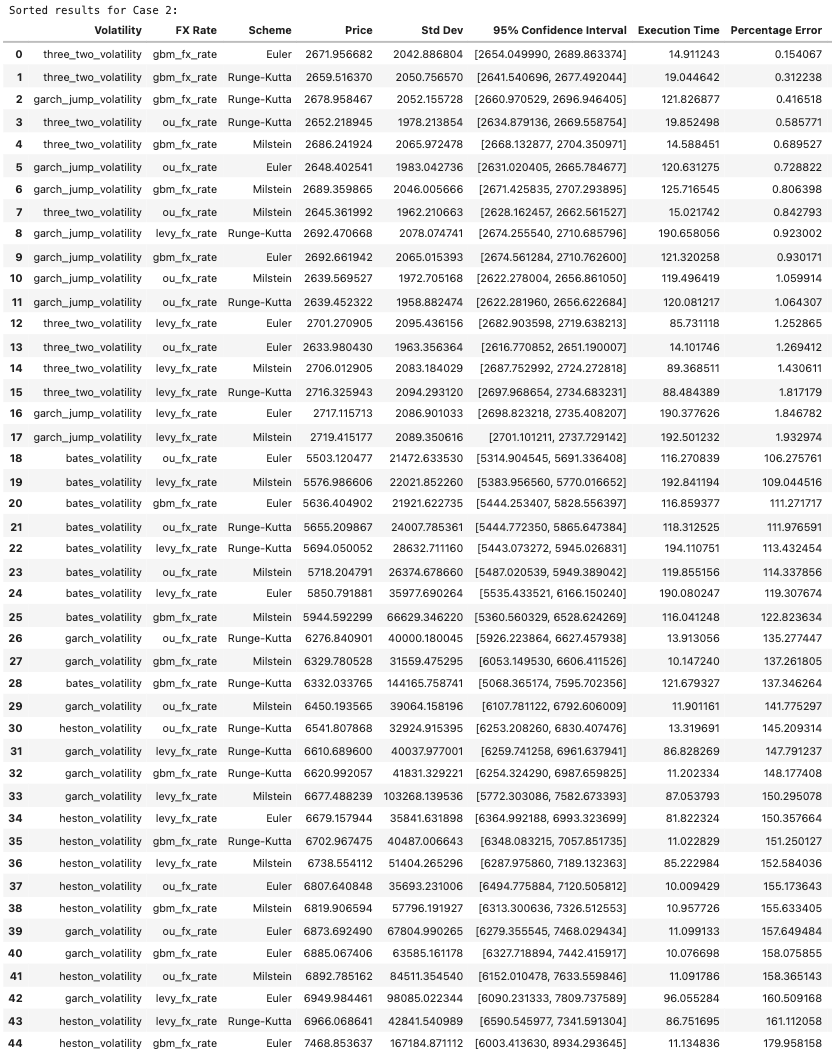} 
    \caption{Performance of MC Simulation with Constant Correlation by Percentage Error, 2021 Start, Case 2}
    \label{fig:MC_Sim_constant_corr_by_percentage_error_2021_start_case_2}
\end{figure}

\begin{figure}[H]
    \centering
    \includegraphics[width=0.85\linewidth]{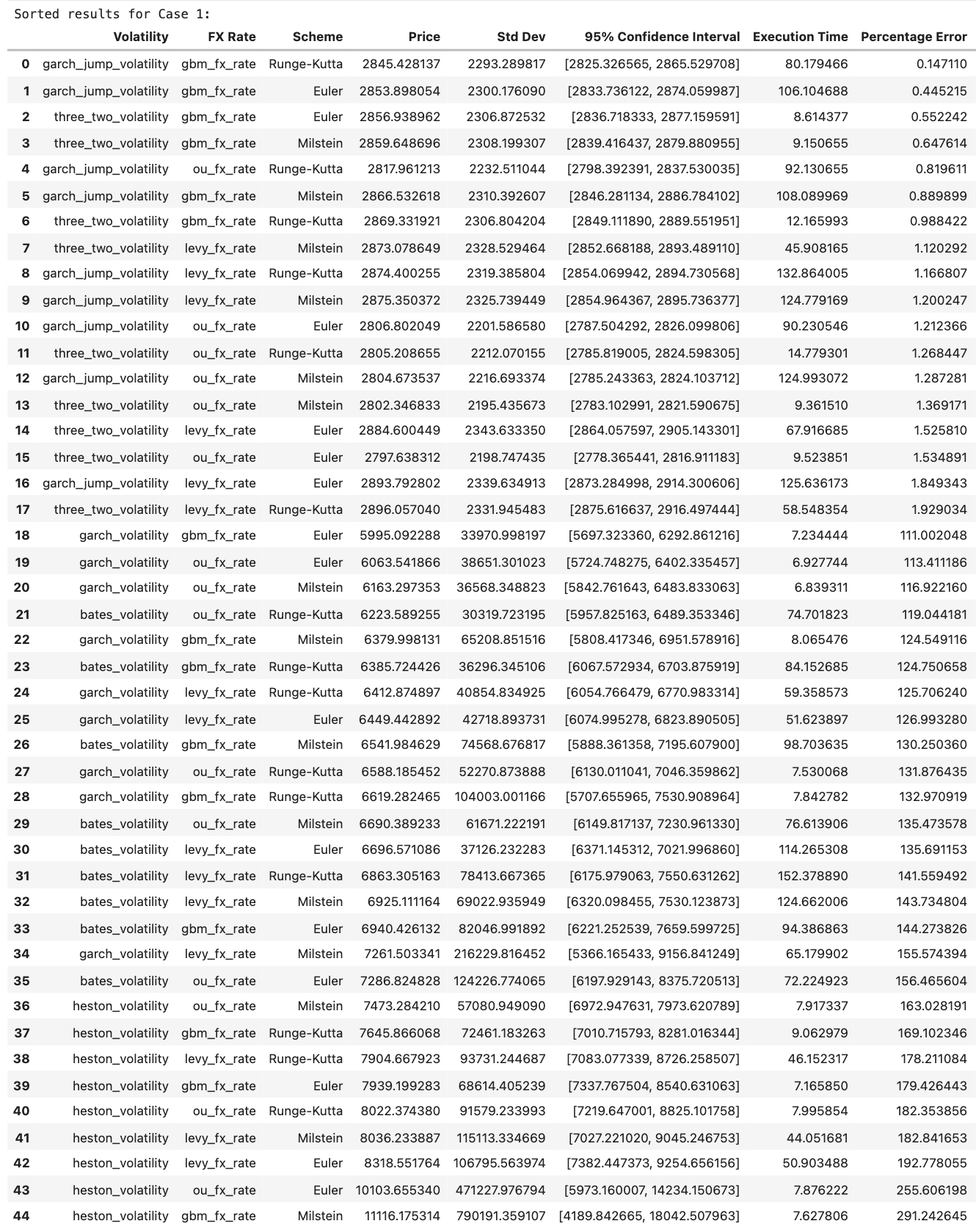} 
    \caption{Performance of MC Simulation with Constant Correlation by Percentage Error, 2022 Start, Case 1}
    \label{fig:MC_Sim_constant_corr_by_percentage_error_2022_start_case_1}
\end{figure}

\begin{figure}[H]
    \centering
    \includegraphics[width=0.85\linewidth]{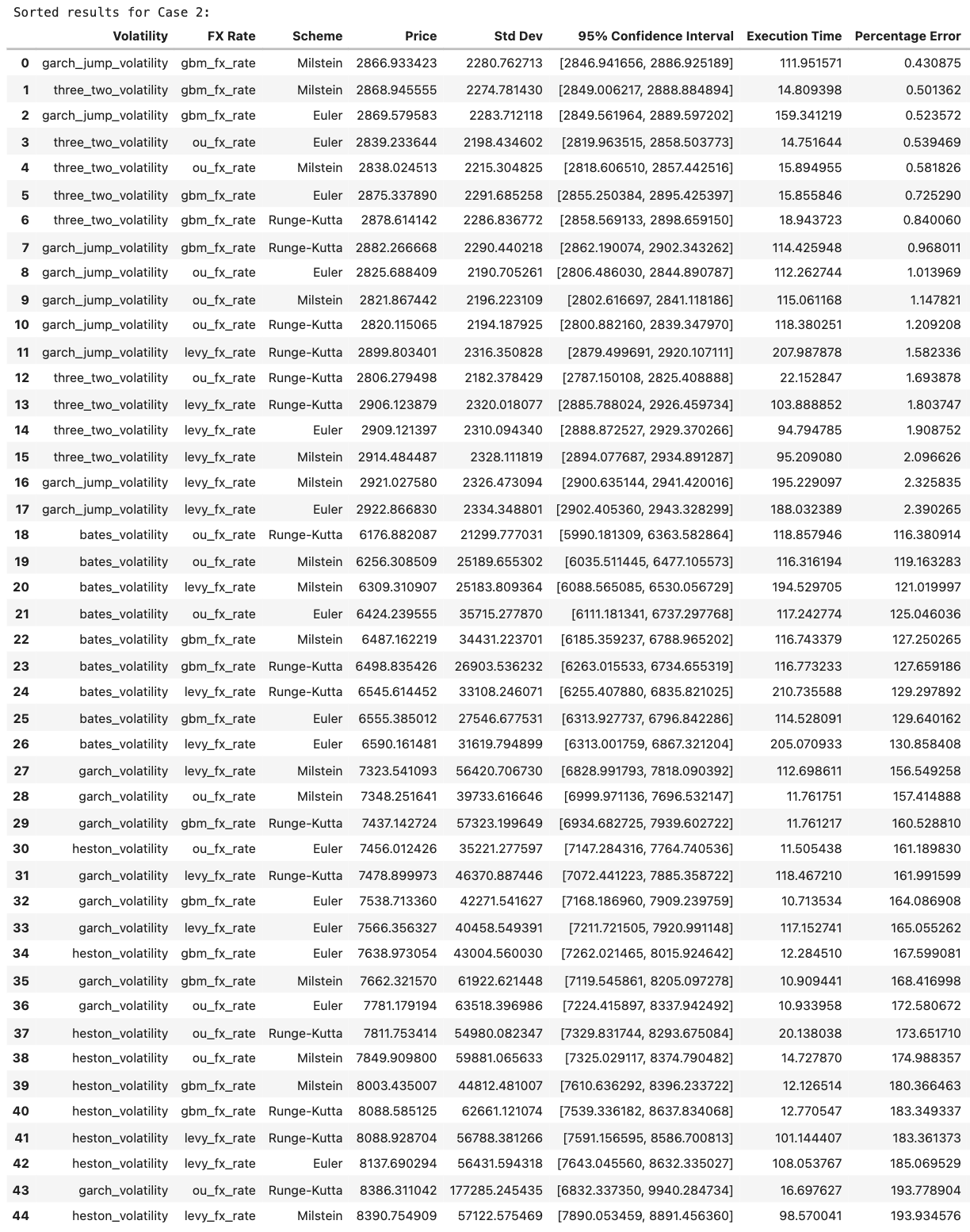} 
    \caption{Performance of MC Simulation with Constant Correlation by Percentage Error, 2022 Start, Case 2}
    \label{fig:MC_Sim_constant_corr_by_percentage_error_2022_start_case_2}
\end{figure}

\begin{figure}[H]
    \centering
    \includegraphics[width=0.95\linewidth]{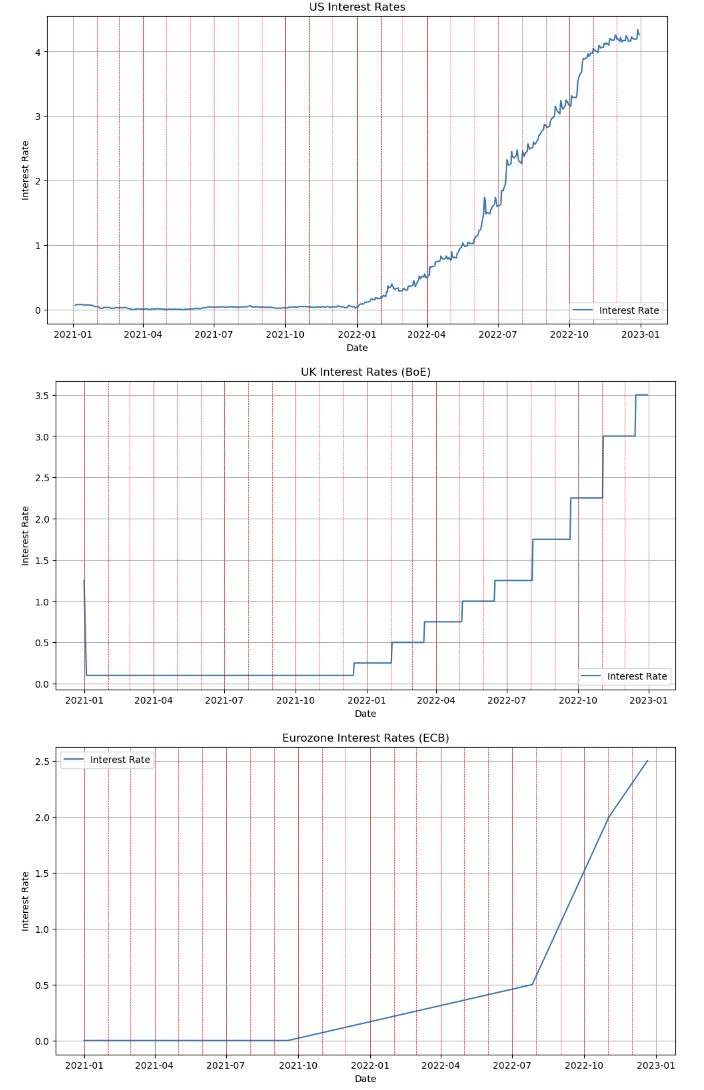} 
    \caption{Interest Rates}
    \label{fig:IRs}
\end{figure}

\begin{figure}[H]
    \centering
    \includegraphics[width=0.65\linewidth]{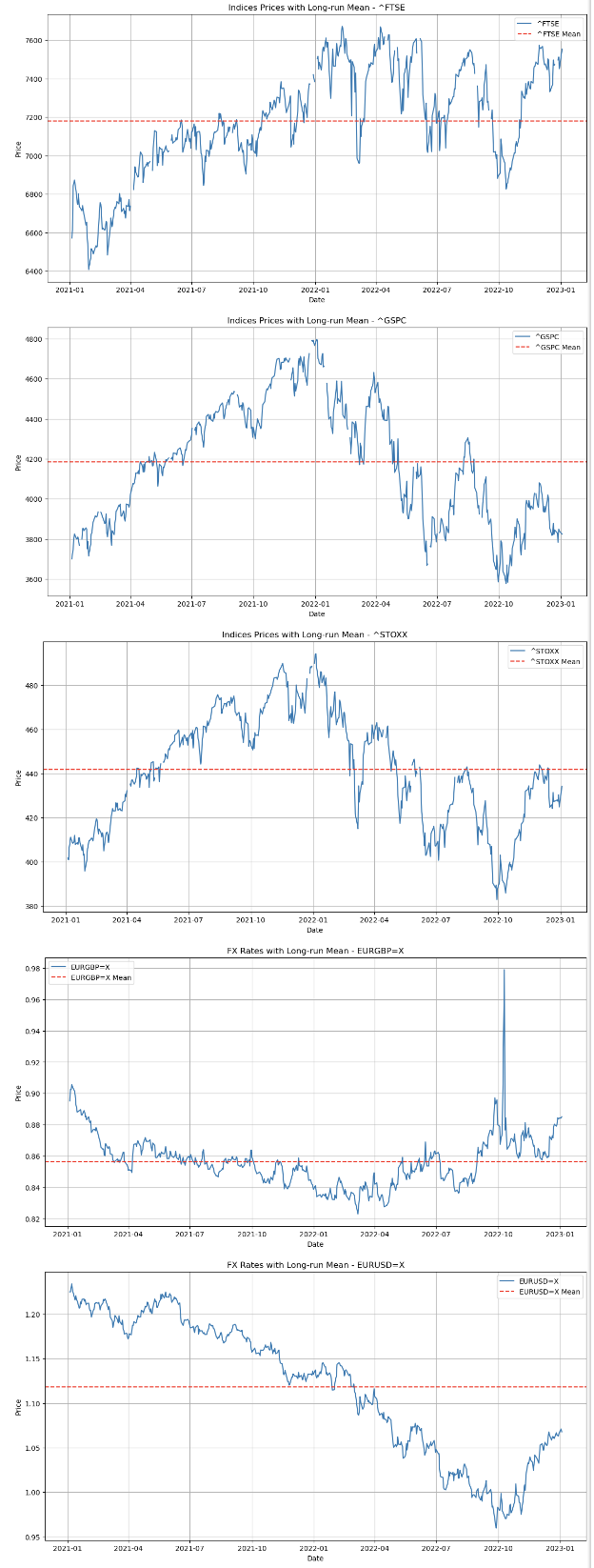} 
    \caption{Plots of Observed Underlying Asset Prices and Exchange Rates 1/2}
    \label{fig:Observed_prices_1}
\end{figure}

\begin{figure}[H]
    \centering
    \includegraphics[width=0.85\linewidth]{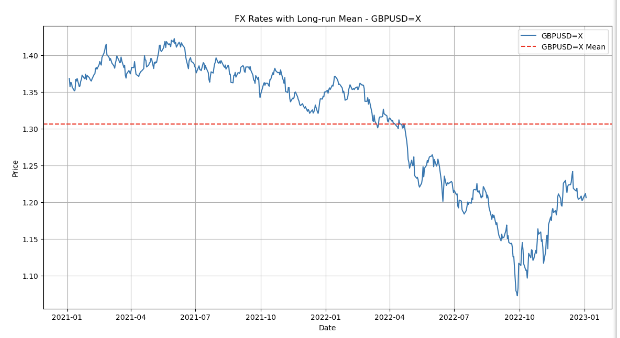} 
    \caption{Plots of Observed Underlying Asset Prices and Exchange Rates 2/2}
    \label{fig:Observed_prices_2}
\end{figure}

\begin{figure}[H]
    \centering
    \includegraphics[width=0.65\linewidth]{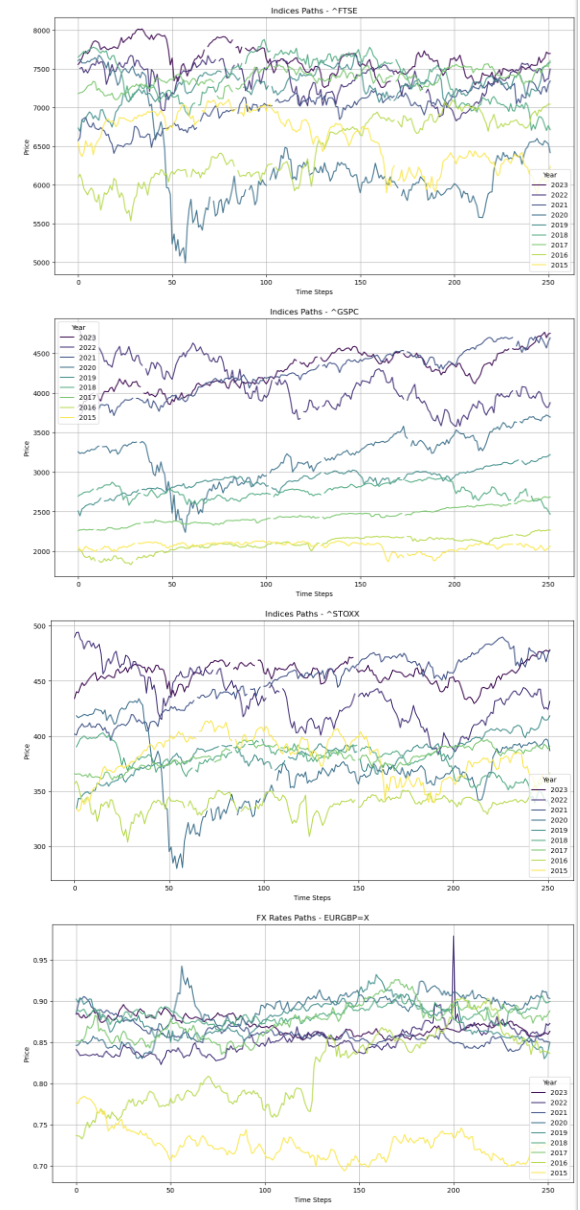} 
    \caption{Plot of Observed Underlying Asset Prices and Exchange Rates in Multiple Years}
    \label{fig:Observed_prices_multiple_years}
\end{figure}

\begin{figure}[H]
    \centering
    \includegraphics[width=0.85\linewidth]{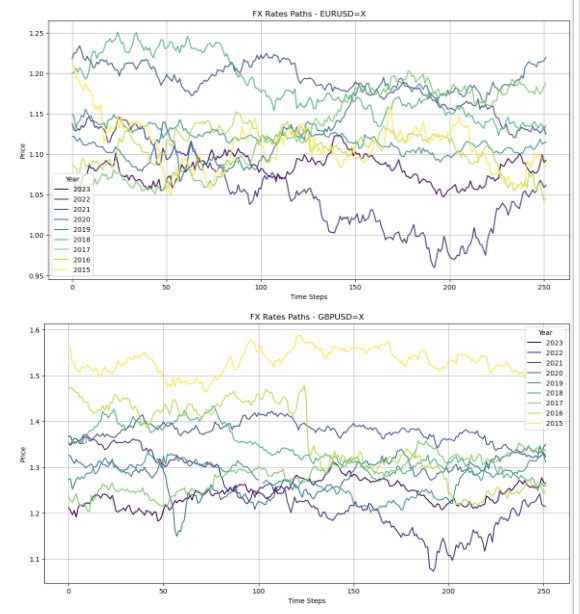} 
    \caption{Plot 2 of Observed Underlying Asset Prices and Exchange Rates in Multiple Years}
    \label{fig:Observed_prices_multiple_years_2}
\end{figure}

\begin{figure}[H]
    \centering
    \includegraphics[width=0.65\linewidth]{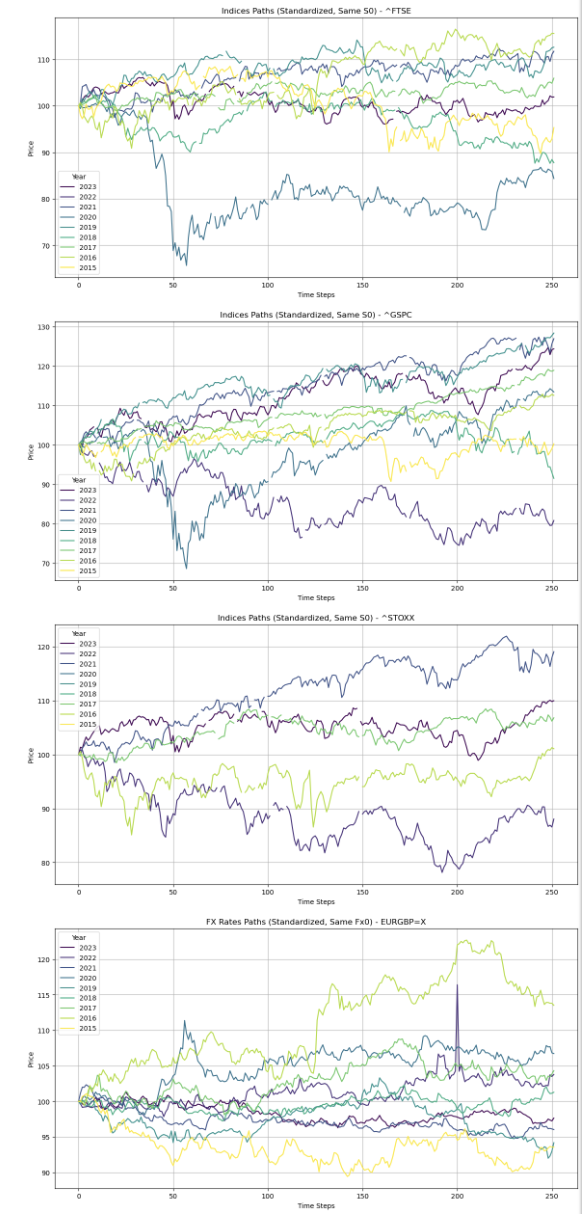} 
    \caption{Plot of Observed Underlying Asset Prices and Exchange Rates in Multiple Years w. Aligned Start}
    \label{fig:Observed_prices_multiple_years_aligned}
\end{figure}

\begin{figure}[H]
    \centering
    \includegraphics[width=0.85\linewidth]{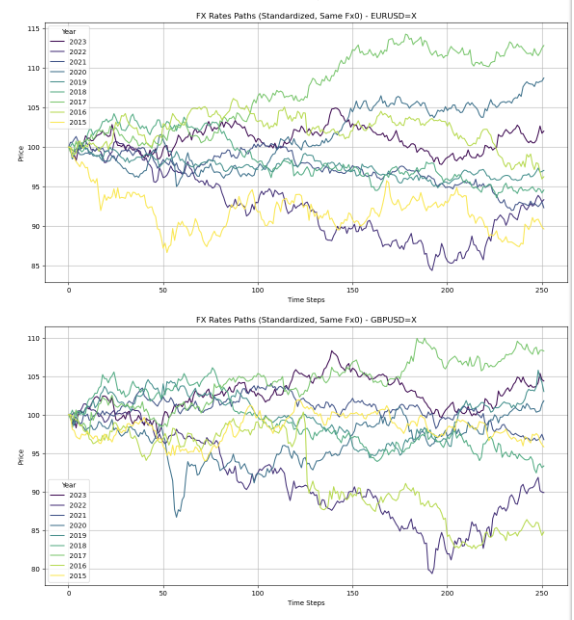} 
    \caption{Plot 2 of Observed Underlying Asset Prices and Exchange Rates in Multiple Years w. Aligned Start}
    \label{fig:Observed_prices_multiple_years_aligned_2}
\end{figure}

\begin{figure}[H]
    \centering
    \includegraphics[width=0.65\linewidth]{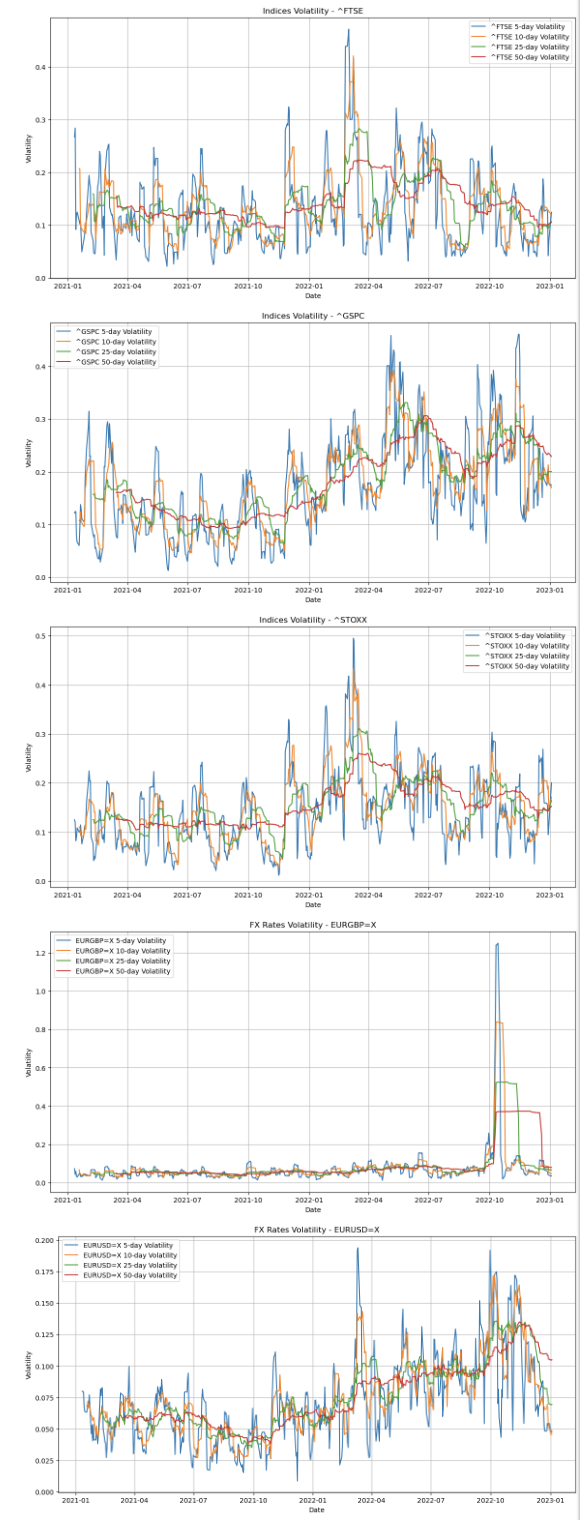} 
    \caption{Observed Volatilities and Correlations 1/2}
    \label{fig:Observed_vol_and_corr}
\end{figure}

\begin{figure}[H]
    \centering
    \includegraphics[width=0.65\linewidth]{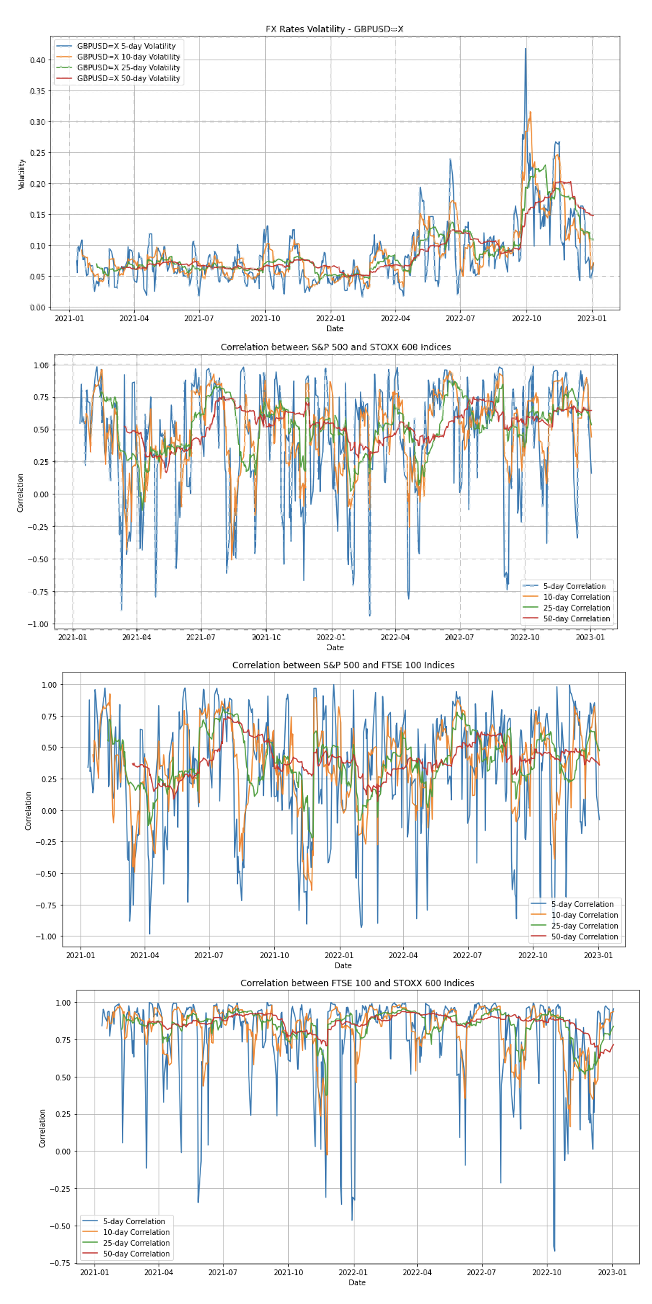} 
    \caption{Observed Volatilities and Correlations 2/2}
    \label{fig:Observed_vol_and_corr_2}
\end{figure}

\begin{figure}[H]
    \centering
    \includegraphics[width=0.59\linewidth]{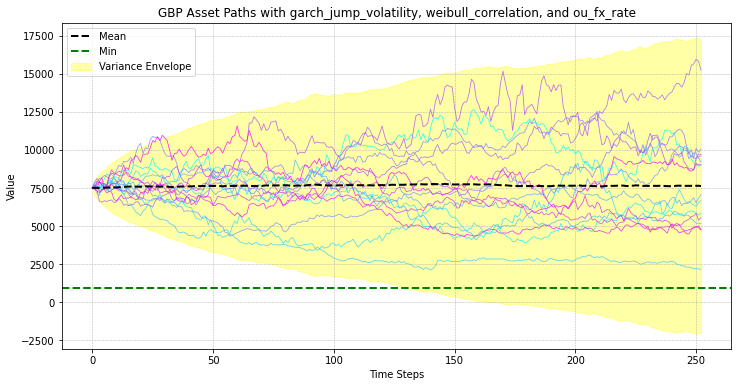} 
    \caption{MC Simulation Paths 1/9}
    \label{fig:Simulation_paths_1}
\end{figure}

\begin{figure}[H]
    \centering
    \includegraphics[width=0.65\linewidth]{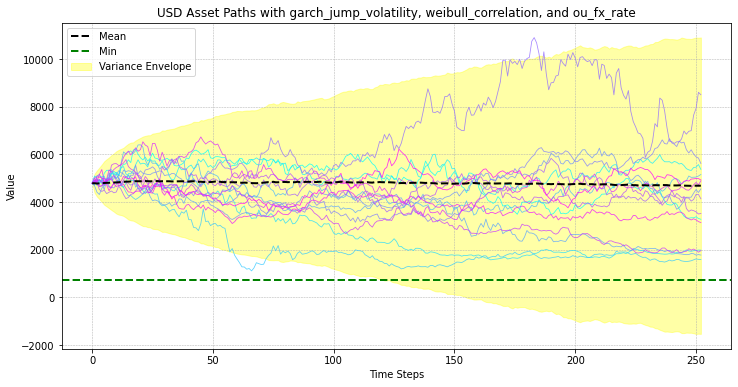} 
    \caption{MC Simulation Paths 2/9}
    \label{fig:Simulation_paths_2}
\end{figure}

\begin{figure}[H]
    \centering
    \includegraphics[width=0.75\linewidth]{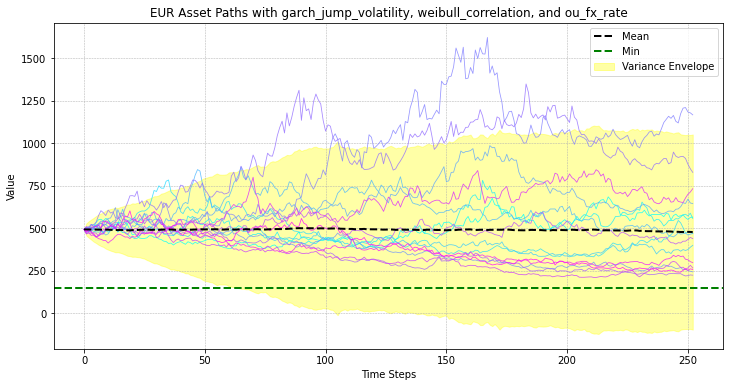} 
    \caption{MC Simulation Paths 3/9}
    \label{fig:Simulation_paths_3}
\end{figure}

\begin{figure}[H]
    \centering
    \includegraphics[width=0.95\linewidth]{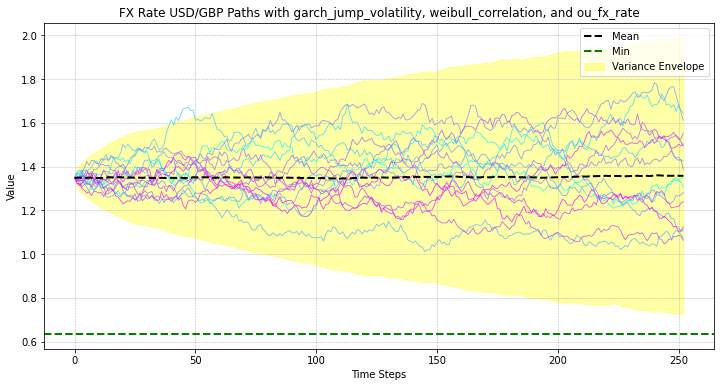} 
    \caption{MC Simulation Paths 4/9}
    \label{fig:Simulation_paths_4}
\end{figure}

\begin{figure}[H]
    \centering
    \includegraphics[width=0.55\linewidth]{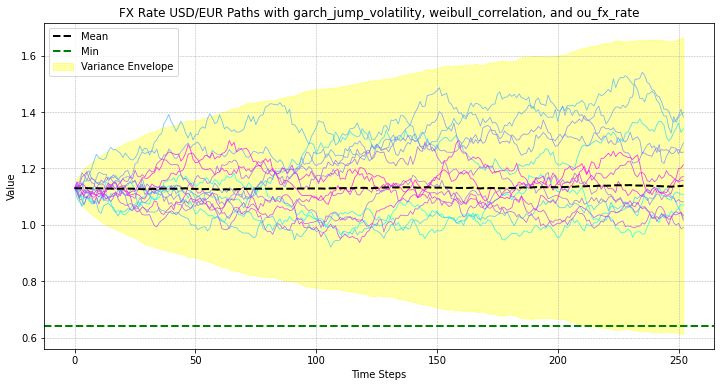} 
    \caption{MC Simulation Paths 5/9}
    \label{fig:Simulation_paths_5}
\end{figure}

\begin{figure}[H]
    \centering
    \includegraphics[width=0.55\linewidth]{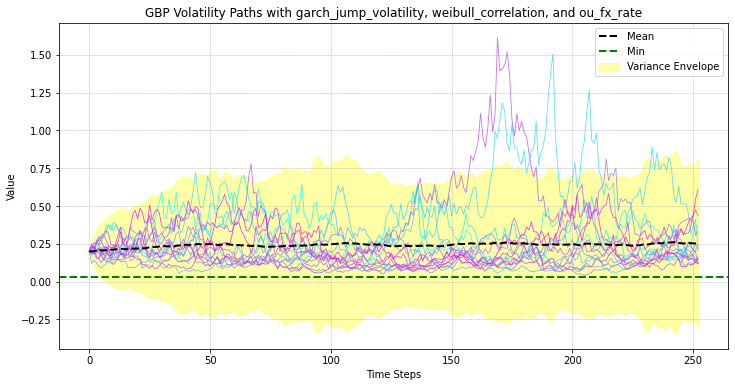} 
    \caption{MC Simulation Paths 6/9}
    \label{fig:Simulation_paths_6}
\end{figure}

\begin{figure}[H]
    \centering
    \includegraphics[width=0.65\linewidth]{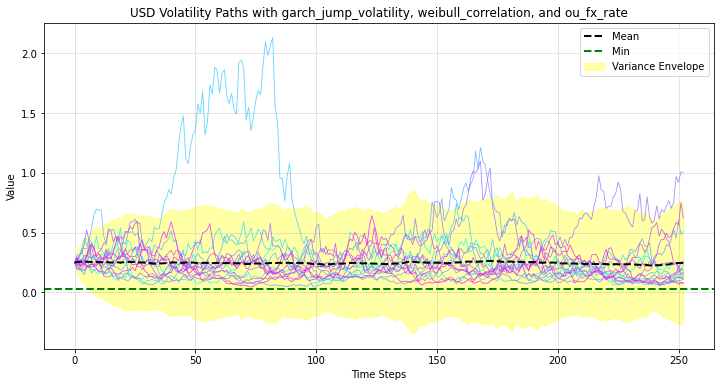} 
    \caption{MC Simulation Paths 7/9}
    \label{fig:Simulation_paths_7}
\end{figure}

\begin{figure}[H]
    \centering
    \includegraphics[width=0.55\linewidth]{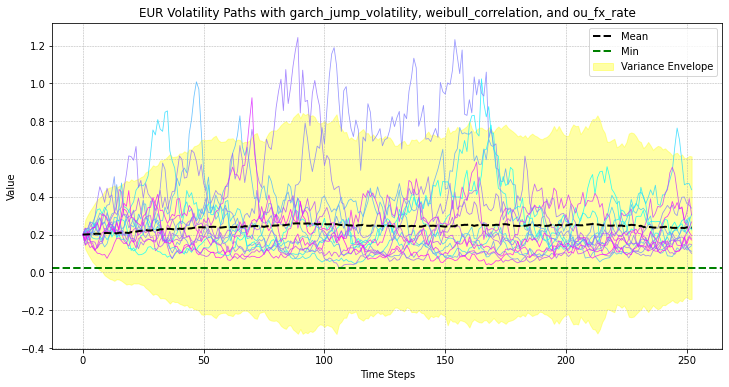} 
    \caption{MC Simulation Paths 8/9}
    \label{fig:Simulation_paths_8}
\end{figure}

\begin{figure}[H]
    \centering
    \includegraphics[width=0.55\linewidth]{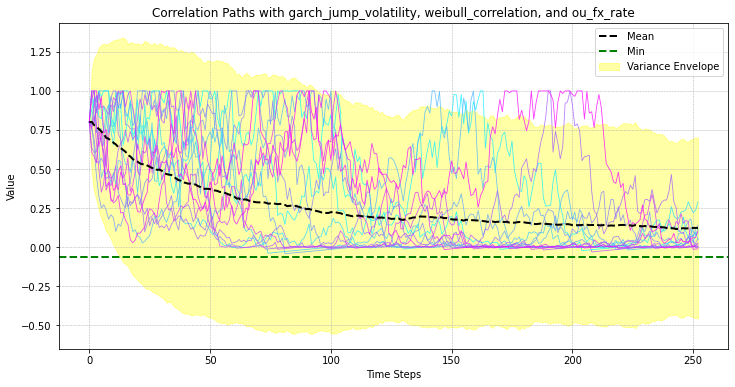} 
    \caption{MC Simulation Paths 9/9}
    \label{fig:Simulation_paths_9}
\end{figure}

\begin{figure}[H]
    \centering
    \includegraphics[width=0.80\linewidth]{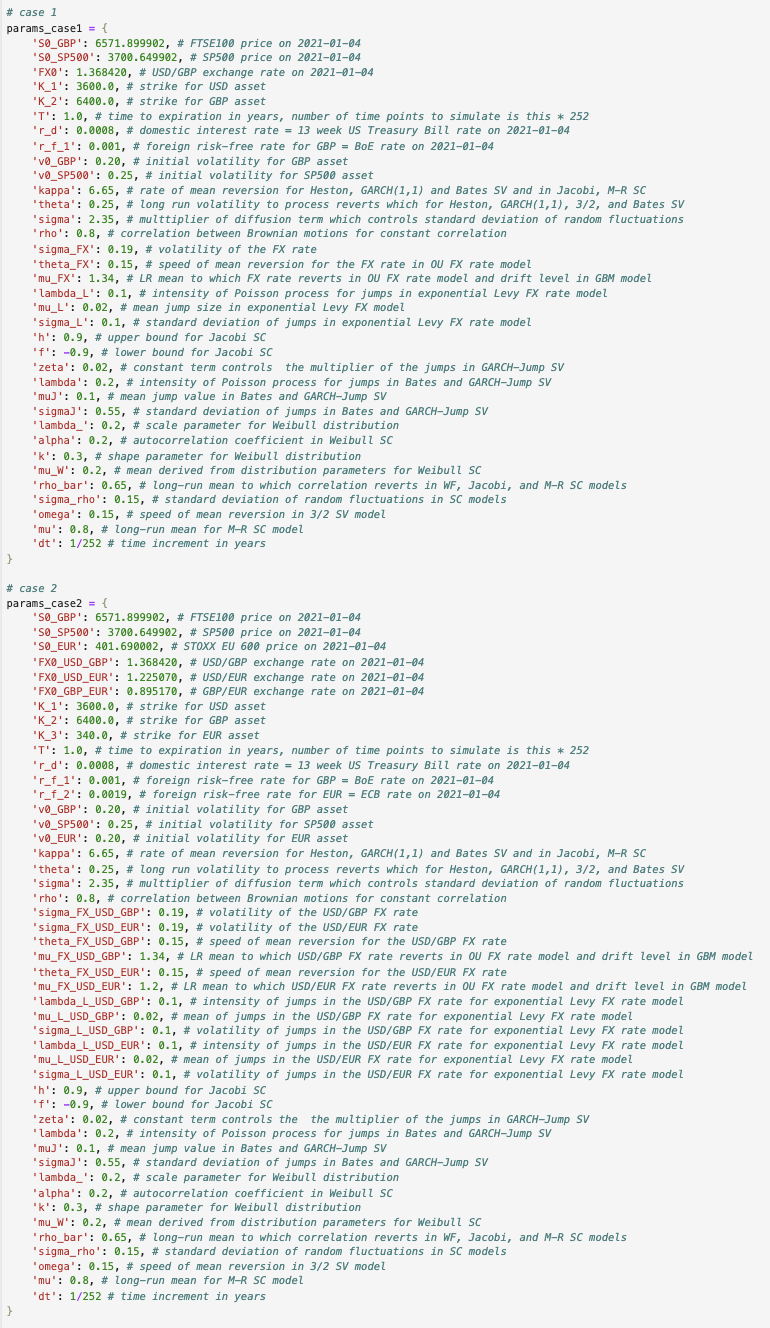} 
    \caption{MC Simulation Parameters for 2021-2022}
    \label{fig:MC_Sim_params_start_2021}
\end{figure}

\begin{figure}[H]
    \centering
    \includegraphics[width=0.80\linewidth]{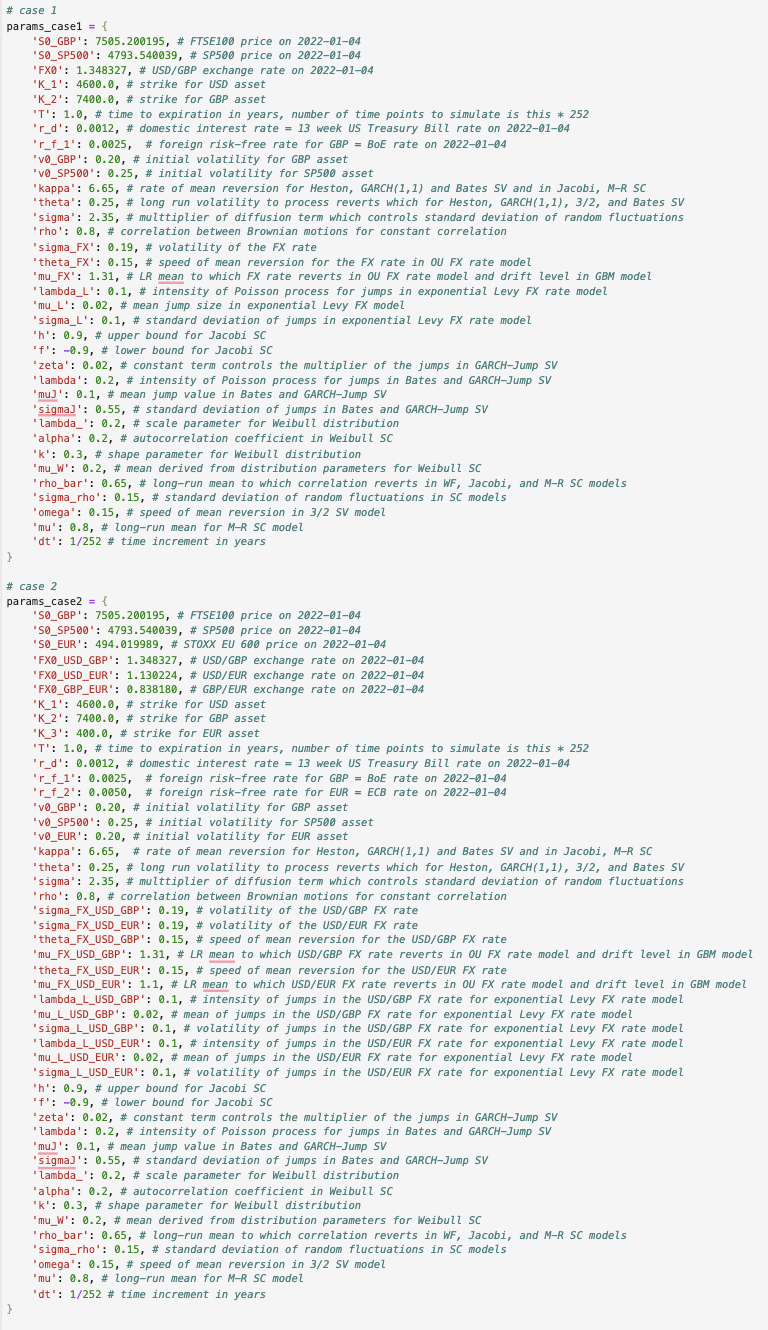} 
    \caption{MC Simulation Parameters for 2022-2023}
    \label{fig:MC_Sim_params_start_2022}
\end{figure}

\end{document}